\documentclass[lettersize,journal]{IEEEtran}
\usepackage{amsmath,amsfonts}
\usepackage{algorithmic}
\usepackage{algorithm}
\usepackage{array}
\usepackage{textcomp}
\usepackage{stfloats}
\usepackage{url}
\usepackage{verbatim}
\usepackage{graphicx}
\usepackage{cite}
\usepackage{subfloat}
\usepackage{subfigure}
\usepackage{color,soul}
\usepackage{bm}
\usepackage{epstopdf}
\usepackage{extarrows}
\makeatletter

\newcommand{\Rmnum}[1]{\expandafter\@slowromancap\romannumeral #1@}
\makeatother
\usepackage{xcolor}
\usepackage{xpatch}

\usepackage{siunitx}
\usepackage{amsthm}          % 放在 amsmath 之后
\begin{document}
	\title{Exploiting Self-Sustainable Information-Bearing RIS in Underlay CR-NOMA Networks}
	\author{Zeyang~Sun,~\IEEEmembership{Graduate~Student~Member,~IEEE},
		Shuai~Han,~\IEEEmembership{Senior~Member,~IEEE},
		Chenyu~Wu,~\IEEEmembership{Member,~IEEE,}
		Sai~Xu,~\IEEEmembership{Member,~IEEE},
		and Yuanwei~Liu,~\IEEEmembership{Fellow,~IEEE}
		\thanks{This work was supported in part by the National Key Research and Development Program of China under Grant 62401175, in part by the National Natural Science Foundation of China under Grants 62431009 and 62371166, and in part by the China Postdoctoral Science Foundation under Grant 2024M764187. Part of this paper has been submitted to the 2026 IEEE ICCC. (\textit{Corresponding author: Chenyu Wu.})}
		\thanks{Z. Sun, S. Han and Chenyu Wu are with the School of Electronics and Information Engineering, Harbin Institute of Technology, Harbin 150001, China, and also with China-Chile ICT “Belt and Road” Joint Laboratory, Harbin 150001, China (e-mail: zeyangsun97@gmail.com; hanshuai@hit.edu.cn; wuchenyu@hit.edu.cn).}
		\thanks{S. Xu is with the Department of Electronic and Electrical Engineering, University College London, London WC1E 6BT, U.K. (e-mail: sai.xu@ucl.ac.uk).}
		\thanks{Y. Liu is with the Department of Electrical and Electronic Engineering, The University of Hong Kong, Hong Kong (e-mail: yuanwei@hku.hk).}
		
		}
	
	\maketitle 
	\begin{abstract}
Information-bearing reconfigurable intelligent surfaces (IB-RIS) provide a promising solution to self-sustainable and green communications by harvesting ambient radio frequency energy while embedding information via passive reflection. This paper investigates a self-sustainable IB-RIS (SIB-RIS)-assisted non-orthogonal multiple access (NOMA) network operating in an underlay cognitive radio (CR) system. Specifically, a multi-antenna primary transmitter (PT) serves a primary user (PU) and concurrently illuminates the secondary nodes, which enables each SIB-RIS to perform simultaneous energy harvesting and backscatter-based information embedding at each RIS. Based on this model, a weighted sum spectral efficiency (WSSE) maximization problem is formulated for the secondary network by jointly optimizing the PT transmit beamforming vector, the SIB-RIS reflection coefficients, and the power-splitting ratios. To tackle the intricately-coupled non-convex problem, an efficient block coordinate descent (BCD) optimization framework is developed, which leverages fractional programming via Lagrangian dual and quadratic transforms together with a difference-of-convex programming approach. Numerical results demonstrate that the proposed SIB-RIS-assisted NOMA CR system yields substantial WSSE gains over both orthogonal multiple access (OMA)-based and active antenna schemes. Moreover, a 2-bit discrete-phase SIB-RIS implementation achieves competitive to which WSSE performance, confirming the practicality of the low-resolution architecture.
	\end{abstract} 
	

	\section{Introduction}
	The rapid proliferation of the Internet of Things (IoT) and the emergence of data-driven applications in sixth-generation (6G) wireless networks are envisioned to drive an unprecedented increase in device density and traffic volume \cite{R1,R2}. On the one hand, licensed spectrum is becoming increasingly congested, and conventional orthogonal multiple access (OMA) schemes assign time-frequency resources to different users in a strictly non-overlapping manner, making it difficult to fully exploit the available bandwidth and thus restricting the number of users that can be served simultaneously \cite{R3}. On the other hand, most existing communication architectures rely on active radio frequency (RF) chains at information transmitters, where signal generation, up-conversion, and power amplification incur non-negligible hardware cost and power consumption at the source side, which conflicts with the vision of green and sustainable 6G networks \cite{R4,Add2}. These phenomena motivate the development of new transmission paradigms that can improve spectrum utilization and reduce the dependence on power-hungry active RF front-ends.

    Cognitive radio (CR) and non-orthogonal multiple access (NOMA) have been recognized as two enabling technologies for alleviating spectrum scarcity and supporting massive connectivity. In underlay CR networks, a secondary network (SN) is permitted to reuse the licensed spectrum of a primary network (PN), provided that the quality-of-service (QoS) requirement of the primary user (PU) is guaranteed \cite{R5}. Such spectrum reuse can substantially improve spectrum utilization without requiring additional dedicated bandwidth. Meanwhile, NOMA allows multiple users to access the same time-frequency resource block via power-domain superposition, while successive interference cancellation (SIC) is employed at the receiver to separate the superimposed signals, thereby enhancing spectral efficiency (SE) and user connectivity compared with OMA schemes \cite{R6}. Despite these advantages, the achievable performance gains of CR and NOMA heavily rely on wireless propagation conditions \cite{R7,R8,R80}. In particular, the satisfaction of the PU's QoS constraint in underlay CR and the assurance of SIC feasibility in NOMA are both governed by the effective channel strengths and their separability. However, for conventional transceivers, the propagation environment is largely uncontrollable, which may severely limit the attainable performance improvements.
    
    In this context, reconfigurable intelligent surfaces (RISs) have recently attracted considerable attention as a low-power paradigm for enabling a programmable propagation environment by adapting the reflection coefficients of reflective elements in a cost-effective manner \cite{R9}. Such controllable propagation is well suited to underlay CR systems with PU QoS constraints, since RIS configuration can simultaneously enhance the desired secondary links and mitigate the interference leakage toward the PU, thereby enlarging the feasible spectrum-sharing region. Moreover, RISs offer additional spatial degrees of freedom (DoFs) to reshape the composite channels of multiple users, which can improve user separability and facilitate reliable SIC in power-domain NOMA. Nevertheless, in most existing RIS-aided CR and NOMA studies \cite{R7,R8,R10,R11}, RIS is primarily leveraged as a purely passive reflector that assists transmission, while the SN still relies on conventional active radio front-ends to generate information-bearing waveforms. This motivates the exploration of new information transmission mechanisms that leverage external RF signals, which in turn reduce the dependence on power-hungry active RF circuitry.
    
    On a parallel track, backscatter communication (BackCom) has emerged as an ultra-low-power transmission paradigm particularly suitable for battery-limited IoT devices \cite{R13}. In BackCom systems, a backscatter device (BD) does not actively generate an RF waveform; instead, it conveys information by modulating and reflecting an incident carrier emitted by an external RF source. This architecture eliminates power-hungry components such as mixers, local oscillators, and power amplifiers \cite{R14}, thereby significantly reducing hardware cost and energy consumption. Recent studies have further shown that BackCom can be naturally integrated with CR \cite{R15,R16}, where BDs in the SN reuse the primary signal as the carrier and superimpose their information through reflection to enable low-cost spectrum sharing and dense connectivity. Nevertheless, in conventional CR-BackCom systems \cite{R17,R18,R19,R20}, BDs are typically implemented as small-aperture tags with limited reflection gain and no beamforming capability, which fundamentally constrains the achievable communication performance. This calls for large-aperture, programmable reflective structures that can simultaneously provide beamforming gain while enabling information embedding through reflection.
    	
    To overcome this limitation, information-bearing RIS (IB-RIS) has been proposed as an extension of conventional RIS architectures \cite{Add1,R21,R22}. Specifically, by appropriately controlling their reflection coefficients, IB-RIS can be exploited not only for channel reconfiguration but also for conveying information, without having to add additional active RF chains at the transmit side \cite{R23,R24}. Compared with conventional BDs, an IB-RIS offers array and beamforming gains for the backscattered signal and additional spatial DoFs for controlling the propagation environment. Recent advances on IB-RIS have significantly extended its functionality through diverse architectural designs. Zhao \emph{et al.} \cite{R25} analytically characterized the probability that a RIS-assisted backscatter link outperforms the direct link and established the fundamental propagation limits for reliable transmission. Tang \emph{et al.} \cite{R26} designed a metasurface-based transmitter capable of joint amplitude-phase modulation, enabling RIS-based multiple-input multiple-output (MIMO) quadrature amplitude modulation (QAM) transmission under practical hardware constraints. Zhao \emph{et al.} \cite{R27} introduced the RIScatter framework, where reflection-state distributions are optimized to jointly encode information and engineer the wireless channel. Sanila \emph{et al.} developed an active RIS architecture that employs joint spatial and reflecting modulation to mitigate multiplicative fading in RIS-assisted MIMO links \cite{R28}, while Xu \emph{et al.} investigated RIS-backscatter-enabled downlink multi-cell MIMO networks, in which RISs operate as passive base stations and element clustering is used to reduce implementation complexity \cite{R29}. 
	
%	\subsection{Motivations and Contributions}
	 Despite these advantages, existing IB-RIS designs still exhibit several limitations from the system perspective. First, most of these works focus on single-link transmission or rely on spatial-division multiple access (SDMA) to separate users in the spatial domain. Such schemes necessitate the deployment of multiple RF chains and rely on favorable channel conditions. Moreover, they fail to exploit the multiplexing capability of NOMA. Second, most designs assume a dedicated RF source or power beacon to illuminate the RIS for backscatter transmission, which still occupies extra spectrum resources for the carrier signal and entails additional costs. Third, the energy consumption of the RIS controller is usually ignored or simply satisfied by external power supplies or batteries. Therefore the overall energy efficiency and the self-sustainability of IB-RIS operation are not explicitly addressed. 
	 
	 To address the above issues, we propose a self-sustainable information-bearing RIS (SIB-RIS)-assisted NOMA CR transmission framework. We aim to maximize the WSSE of the SN while strictly guaranteeing the PU's quality-of-service (QoS), and the resulting resource allocation problem is challenging to tackle due to the intricate coupling among the primary transmitter (PT) beamforming vector, the SIB-RIS reflection coefficients, and the power splitting (PS) ratios. \emph{To the best of our knowledge, the joint design of SIB-RIS and NOMA within a CR framework has not been systematically investigated}. The main contributions are summarized as follows:
	
	\begin{itemize}
		\item We propose a novel underlay CR framework where multiple SIB-RISs are deployed at the secondary transmitters (STs). Each SIB-RIS adopts a PS structure to harvest energy from the primary signal and utilizes the remaining power to perform information-bearing backscatter modulation. Therefore, the SIB-RISs serve as self-sustainable passive transmitters, enabling uplink NOMA transmission from the SN to an access point (AP).
		\item Based on the established framework, we formulate a WSSE maximization problem for the SN, subject to the PT transmit power budget, PU QoS constraint, SIB-RIS energy harvesting constraint, and unit-modulus reflection constraint. Due to the highly-coupled and non-convex structure of the problem, we develop an efficient block coordinate descent (BCD) framework that leverages fractional programming, Lagrangian dual transformation, and difference-of-convex (DC) to jointly optimize the power-splitting coefficients at the SIB-RISs, the active beamforming vector at the PT, and the passive reflection coefficients of SIB-RISs.
		\item Extensive numerical results are provided to evaluate the proposed SIB-RIS-assisted NOMA CR system. The main insights are as follows: 1) The proposed scheme consistently outperforms the SIB-RIS-assisted OMA benchmark in terms of WSSE; 2) Over a broad range of system configurations, the proposed SIB-RIS-assisted NOMA architecture achieves higher WSSE than the conventional active-antenna NOMA baseline; 3) The 2-bit discrete-phase SIB-RIS implementation attains performance comparable to the continuous-phase design while still yielding significant gains over the active-antenna benchmark, which confirms the practical viability of the proposed architecture with low-resolution hardware.
	\end{itemize}
	
	\subsection{Organization and Notation}
	
	The remainder of this paper is organized as follows. Section~II describes the system architecture of the proposed SIB-RIS-assisted NOMA CR network and formulates the WSSE maximization problem. Section~III develops the proposed BCD framework, including the equivalent problem transformation and the joint optimization scheme. Section~IV presents extensive numerical results to validate the effectiveness of the proposed scheme. Finally, Section~V concludes the paper.
	
	\textit{Notation:} Scalars, vectors, and matrices are denoted by italic letters, bold lowercase and bold uppercase letters, respectively. The transpose, conjugate transpose, and inverse of a matrix $\mathbf{X}$ are denoted by $\mathbf{X}^{{T}}$, $\mathbf{X}^{{H}}$, and $\mathbf{X}^{-1}$, respectively. The Euclidean norm of a vector $\mathbf{x}$ is written as $\|\mathbf{x}\|$, and $|x|$ denotes the absolute value of a scalar $x$. The operators $\mathrm{Tr}(\cdot)$, $\mathrm{rank}(\cdot)$, and $\mathrm{diag}(\cdot)$ stand for the trace, rank, and the diagonal-matrix operator, respectively. $\Re\{\cdot\}$ denotes the real part of a complex quantity, and $[ \mathbf{X} ]_{k,k}$ denotes the $(k,k)$-th diagonal element of $\mathbf{X}$. The notation $\mathbf{X} \succeq \mathbf{0}$ indicates that $\mathbf{X}$ is a Hermitian positive semidefinite matrix. The complex Gaussian distribution with mean $\boldsymbol{\mu}$ and covariance matrix $\boldsymbol{\Sigma}$ is denoted by $\mathcal{CN}(\boldsymbol{\mu},\boldsymbol{\Sigma})$, and $\mathbb{E}\{\cdot\}$ denotes statistical expectation. Finally, $\mathbb{C}^{m \times n}$ represents the set of all $m \times n$ complex-valued matrices.

	\section{System Architecture and Problem Formulation}	
	\subsection{System Architecture}
	As illustrated in Fig.~1, we consider a narrowband underlay CR system, where a PN coexists with a SN enabled by multiple SIB-RISs. Specifically, the PN consists of a PT equipped with $N$ antennas and a single-antenna PU. The PT broadcasts a primary signal $c$ with normalized power, i.e., $\mathbb{E}\left\{|c|^2\right\} = 1$, which simultaneously serves as both the energy source and the carrier for the SIB-RIS. The SN comprises $M$ STs and an AP. Each ST is integrated with a $K$-element SIB-RIS that employs a power-splitting front end to enable simultaneous energy harvesting and information-bearing BackCom. Unlike the conventional RISs that merely reflect incident signals without altering their intrinsic properties, the SIB-RISs act as passive transmitters. By embedding secondary information onto the reflected primary signal through backscatter communication and properly designing the SIB-RIS reflection coefficients, SIB-RISs realize simultaneous self-sustainable uplink NOMA transmission toward the AP without requiring dedicated RF components. The AP serves as the receiver for the SN, and performs the SIC to sequentially decode the secondary messages.
	
	We assume that all channels undergo quasi-static fading process and the channel state information (CSI) can be perfectly acquired\footnote{The acquisition of CSI for RIS-assisted communication systems is non-trivial due to the passive nature of RISs and is beyond the scope of this work. Various pilot-aided and cascaded-channel estimation schemes have been developed for RIS-assisted systems and demonstrated to achieve high estimation accuracy \cite{R30,R31}. Moreover, the perfect-CSI assumption adopted here mainly serves to characterize a performance upper bound and to provide a benchmark for more practical robust designs under CSI uncertainty.} \cite{R25,R26,R27,R28,R29}. Furthermore, it is assumed that the SIB-RISs are capable of perfect spectrum sensing, thereby ensuring the QoS of the PU. For ease of illustration, the sets of the SIB-RISs and the reflective elements of the $j$-th SIB-RIS can be expressed as ${\mathcal M} = \{ 1, 2, ..., M \}$ and ${\cal K}_j = \left\{ {1,2,...,K} \right\}$, respectively. Let ${{\bf{h}}_{\rm p}} \in {\mathbb{C}^{N \times 1}}$, ${{\bf{h}}} \in {\mathbb{C}^{N \times 1}}$, ${\bf{F}}_j \in {\mathbb{C}^{K \times N}}$, ${{\bf{g}}_{{\rm{p}}j}} \in {\mathbb{C}^{K \times 1}}$ and ${{\bf{g}}_j} \in {\mathbb{C}^{K \times 1}}$ denote the equivalent channel matrices for the PT to the PU link, the PT to the AP link, the PT to the $j$-th SIB-RIS link, the $j$-th SIB-RIS to the PU link, and the $j$-th SIB-RIS to the AP link, respectively.
	
		\begin{figure}[tpb]
		\centering
		\includegraphics[width=3.5in]{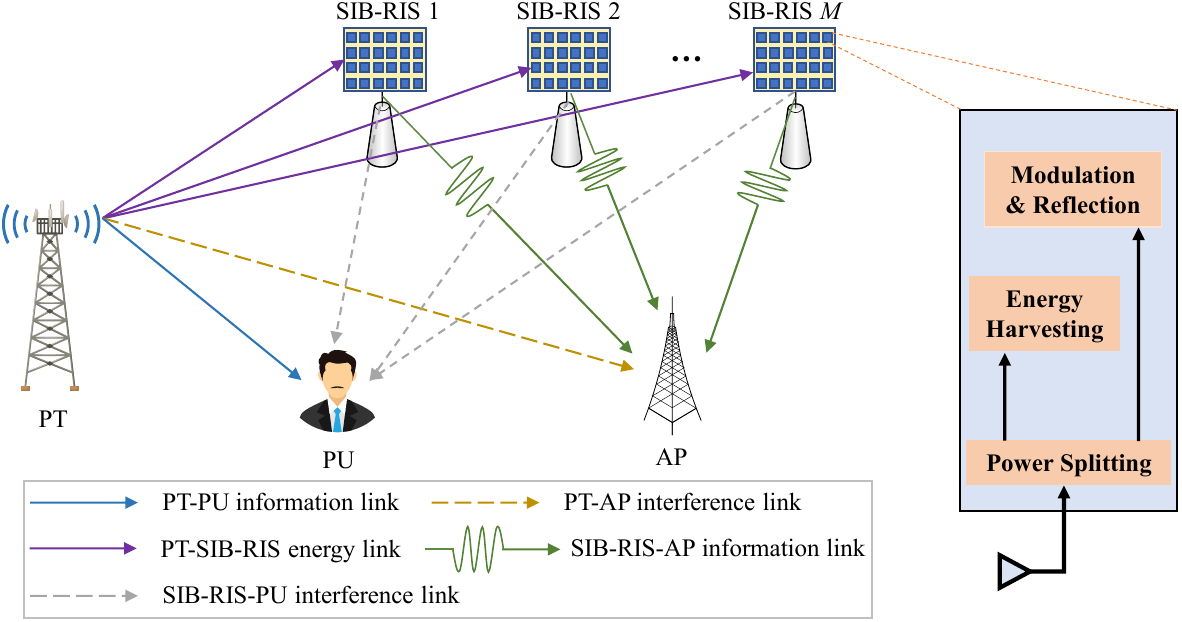}
		\caption{The proposed SIB-RIS-assisted NOMA CR system.}
		\label{Fig1}
	\end{figure}
	
	\subsubsection{Self-Sustainable Power-Splitting Energy Harvesting} In order to achieve self-sustainability, we assume that each SIB-RIS is equipped with an energy harvesting circuit and operates the PS protocol to harvest RF energy from primary signal. Specifically, the incident signal power at $j$-th SIB-RIS is split into two parts: a $1- \delta_j^2$ portion is routed to the energy harvesting circuit, while the remaining power is used for information-bearing backscatter modulation. The corresponding PS coefficient $\delta_j$ satisfying $0 < \delta_j  < 1$.
	
	To simplify the SIB-RIS circuit design and reduce the control circuit power consumption, we assume that all the SIB-RIS elements employ the same PS parameter and the harvested power operates within the linear sensitivity region of the energy harvesting circuit \cite{R32,R33}. The harvested power of the $j$-th SIB-RIS can be denoted as
	\begin{equation}
		P_j^{{\rm{EH}}} = \chi \left( {1 - \delta_j^2 } \right){\left\| {{{\bf{F}}_j}{\bf{w}}} \right\|^2},
		\label{eq1}
	\end{equation}
	where $\chi$ denotes the energy harvesting efficiency and $\bf w$ indicates the active beamforming vector of PT. The energy constraint $P_j^{{\rm{EH}}} \ge \mu K$ must be satisfied to guarantee the self-sustainability of the SIB-RIS, where $\mu$ denotes the energy consumption constant of each SIB-RIS element.

	\subsubsection{Information-Bearing via Backscatter Modulation}
	The information-bearing process is realized via backscatter modulation by
	appropriately adjusting the reflection coefficients of the SIB-RIS elements,
	so that each SIB-RIS can superpose its own information onto the incident
	primary signal. Specifically, the information embedding process at the $j$-th
	SIB-RIS can be mathematically formulated as  
	\begin{equation}
		\begin{aligned}
			\delta_j \mathbf g_j^{ H}\mathbf\Theta_j\mathbf F_j\mathbf w c
			&= \delta_j \mathbf g_j^{ H}
			\mathrm{diag}\!\big(\mathbf F_j\mathbf w\big)\boldsymbol\theta_j c\\
			&\xLongrightarrow[]{\text{information embedding}}
			\delta_j \mathbf g_j^{ H}
			\mathrm{diag}\!\big(\mathbf F_j\mathbf w\big)\boldsymbol\phi_j x_j,
		\end{aligned}
		\label{eq2}
	\end{equation}
	where $\mathbf\Theta_j$ and $\boldsymbol\theta_j$ denote the backscatter
	matrix and vector of the $j$-th SIB-RIS, respectively; $x_j$ denotes the
	transmitted symbol associated with the $j$-th SIB-RIS with
	$\mathbb E\{|x_j|^2\}=1$, and $\boldsymbol\phi_j$ is the corresponding
	passive beamforming vector. The backscatter process is facilitated by
	the transformation from $\boldsymbol\theta_j c$ to $\boldsymbol\phi_j x_j$,
	and the vector $\boldsymbol\theta_j$ can be interpreted as jointly capturing
	the information-bearing modulation and passive beamforming \cite{R23,R24}. Specifically, the backscatter coefficient of the $k$-th element of the
	$j$-th SIB-RIS is decomposed as $[\boldsymbol\theta_j]_k
	= [\boldsymbol\phi_j]_k [\boldsymbol\kappa_j]_k
	= \tau_{j,k} e^{j\psi_{j,k}}\, \xi_{j} e^{j\lambda_j}$, where $\boldsymbol\kappa_j$ denotes the information-bearing vector,
	$\tau_{j,k}\in[0,1]$ and $\xi_{j}\in[0,1]$ are the reflection and modulation
	amplitudes, while $\psi_{j,k}\in[0,2\pi)$ and $\lambda_j\in[0,2\pi)$ are
	the associated phases. In this work, we adopt a constant-modulus
	phase-modulation scheme for the secondary information and set $\xi_{j}=1$.	
	Due to the passive nature of the reflecting elements, the reflection
	amplitudes must satisfy $|[\boldsymbol\phi_j]_k|\le 1$ for all $j\in\mathcal M$ and
	$k\in\mathcal K_j$. To avoid the hardware complexity of amplitude control,
	we further adopt phase-only reflection and impose the unit-modulus
	constraint $[\boldsymbol\phi_j\boldsymbol\phi_j^{H}]_{k,k} = 1,
	\ j\in\mathcal M,~k\in\mathcal K_j$.	
	
\subsubsection{Signal Transmission and NOMA Decoding Model} The received signal at PU can be written as
	\begin{equation}
		\begin{aligned}
			{y_{\rm{p}}} = {\bf{h}}_{\rm{p}}^H{\bf{w}}c + \sum\limits_{i \in {\cal M}} \delta_i {{\bf{g}}_{{\rm p}i}^H{\rm{diag}}\left( {{{\bf{F}}_i}{\bf{w}}} \right){{\bm \phi} _i}{x_i}}  + {z_{\rm{p}}},
		\end{aligned}
		\label{eq3}
	\end{equation}
	where $z_{\rm p} \sim {\cal C}{\cal N}\left( {{{0}},\sigma^2} \right)$ indicates the additive Gaussian white noise at the PU. 
	
	Accordingly, the signal-to-interference-noise ratio (SINR) of the PU is expressed as
	\begin{equation}
		\begin{aligned}
			{\gamma _{\rm{p}}} = \frac{{{{\left| {{{\bf{h}}_{\rm p}^H}{\bf{w}}} \right|}^2}}}{{\sum\limits_{i = 1}^M {{{\left| {{\delta _i}{\bf{g}}_{{\rm p}i}^H{\rm{diag}}\left( {{{\bf{F}}_i}{\bf{w}}} \right){{\bm \phi} _i}} \right|}^2}}  + {\sigma ^2}}},
		\end{aligned}
		\label{eq4}
	\end{equation}
	and the corresponding rate of PU is given by ${R_{\rm p}} = {\log _2}\left( {1 + {\gamma _{\rm p}}} \right)$.

	Similarly, the received signal at the AP can be formulated as  
	\begin{equation}
		\begin{aligned}
			y = {{\bf{h}}^H}{\bf{w}}c + \sum\limits_{i \in {\cal M}} {\delta_i{\bf{g}}_i^H{\rm{diag}}\left( {{{\bf{F}}_i}{\bf{w}}} \right){{\bm \phi} _i}{x_i}}  + {z},
		\end{aligned}
		\label{eq5}
	\end{equation}
	where $z \sim {\cal C}{\cal N}\left( {{{0}},\sigma^2} \right)$ represents the additive Gaussian white noise at the AP. 
	
	In the uplink NOMA communication framework, the SIC technique is employed at the AP to decode superimposed signals through sequential detection process. The SIC protocol prioritizes signal recovery based on effective channel gain hierarchy, where the SIB-RIS node with superior channel conditions is decoded first, while signals from SIB-RIS nodes with relatively weaker channel gains are treated as interference. To formalize this process, we define $\Psi (j)$ as the decoding priority index assigned to the $j$-th SIB-RIS node. When recovering the information symbol $x_j$ conveyed by the $j$-th SIB-RIS node, the AP successively decodes and subtracts signals from all SIB-RIS nodes $n$ satisfying $\Psi (n) < \Psi (j)$ by treating signals from SIB-RIS nodes $i$ with $\Psi (i) > \Psi (j)$ as interference.
	
	The instantaneous cascaded effective gain of the $j$-th SIB-RIS is written as
		$g_j(\mathbf w,\boldsymbol\phi_j)=\left|\delta_j\,\mathbf g_j^{H}\mathrm{diag}(\mathbf F_j\mathbf w)\boldsymbol\phi_j\right|$,
		which is reconfigurable via $\boldsymbol\phi_j$. Since the joint optimization of the SIC decoding
		order and $\{\boldsymbol\phi_j\},\mathbf w$ leads to a mixed-integer nonconvex problem with $M!$
		possible orders, a fixed-order rule is adopted. Specifically, the decoding order is determined once
		according to the reference gain $
			\tilde g_j = \left|\delta_j\,\mathbf g_j^{H}\mathrm{diag}(\mathbf F_j\mathbf w^{(0)})\boldsymbol\phi_{0}\right|$,
		where $\mathbf w^{(0)}$ is initialized as a full-power constant-modulus beamformer with identical real and imaginary parts across antennas and $\boldsymbol\phi_{0}$ is a normalized
		uniform-phase vector. The resulting descending order is denoted by $\Psi(j)=j$ and is kept fixed
		for complexity reduction \cite{R6,R34}. Accordingly, the SINR for decoding the $j$-th SIB-RIS symbol
		at the AP is given by		
%	The effective channel gain of the $j$-th SIB-RIS is given by
%	$\left|\delta_j {\bf{g}}_{j}^H{\rm{diag}}\big( {{{\bf{F}}_j}{\bf{w}}} \big){{\bm \phi} _j} \right|$.
%	However, this gain can be reconfigured via the SIB-RIS reflection coefficients. Consequently, we adopt $\left| {{\bf{g}}_{_j}^H{\rm{diag}}\left( {{{\bf{F}}_j}{\bf{w}}} \right){{\bm \phi} _0}} \right|$ as the initial effective channel gain of $j$-th SIB-RIS, where ${{{\bm \phi} _0}}$ denotes a normalized reference vector with equal real and imaginary components for all entries. Without loss of generality, we further assume the channel gains are sorted based on the initial effective channel gain \cite{R6,R34}. Then the SIB-RISs are ordered in descending order, denoted by $\Psi (j) = j$. Hence, the SINR for the AP to recover the information symbol of $j$-th SIB-RIS is given by
	\begin{equation}
		\begin{aligned}
			{\gamma _j} = \frac{{{{\left| {\delta_j{\bf{g}}_j^H{\rm{diag}}\left( {{{\bf{F}}_j}{\bf{w}}} \right){{\bm \phi} _j}} \right|}^2}}}{{\sum\limits_{i > j}^M {{{\left| {\delta_i{\bf{g}}_i^H{\rm{diag}}\left( {{{\bf{F}}_i}{\bf{w}}} \right){{\bm \phi} _i}} \right|}^2}}  + {{\left| {{\bf{h}}^H{\bf{w}}} \right|}^2} + {\sigma ^2}}},
		\end{aligned}
		\label{eq6}
	\end{equation}
	and its corresponding achievable rate is ${R_j} = {\log _2}\left( {1 + {\gamma _j}} \right)$.
	
	\subsection{Problem Formulation}
	In this paper, we aim to maximize the WSSE of SN via optimizing the active beamforming vector $\bf w$ of the PT, the coefficient vectors ${\bm \phi}_j, j \in \mathcal M$ and the power splitting parameter $\delta_j, j \in \mathcal M$ for the SIB-RISs. Based on the established system model, the WSSE maximization problem can be mathematically formulated as
	\begin{subequations}
		\begin{align}
			\left(\mathrm{P1}\right)\ 
			\max_{\mathbf{w},\mathcal{R},\boldsymbol{\delta}}
			& \sum_{j=1}^{M} \omega_j \log_2\!\left(1+\gamma_j\right),
			\label{eq7a}\\
			\mathrm{s.t.}\quad
			& \mathrm{Tr}\!\left(\mathbf{w}\mathbf{w}^H\right) \le P,
			\label{eq7b}\\
			& \left[\boldsymbol{\phi}_j \boldsymbol{\phi}_j^{H}\right]_{k,k} = 1,
			\; j\in\mathcal{M},\ k\in\mathcal{K}_j,
			\label{eq7c}\\
			& 0 < \delta_j < 1,
			\; j\in\mathcal{M},
			\label{eq7d}\\
			& \chi\!\left(1-\delta_j^2\right)
			\left\|\mathbf{F}_j\mathbf{w}\right\|^2
			\ge \mu K,
			\; j\in\mathcal{M},
			\label{eq7e}\\
			& R_{\mathrm{p}} \ge R^{\mathrm{TH}},
			\label{eq7f}
		\end{align}
	\end{subequations}
	where $\omega_j \ge 0$ denotes the priority weight of the $j$-th SIB-RIS node, $\mathcal R$ represents the collection of ${\bm \phi}_j$, ${\bm \delta} \in \mathbb R^{M \times 1}$ indicates the PS coefficient vector, $P$ is the predefined power budget of the PT, and ${R ^{{\rm{TH}}}}$ denotes the minimum rate requirement for the PU. Constraint (\ref{eq7b}) regulates the maximum transmission power of PT, (\ref{eq7c}) denotes the unit modulus constraint for the $j$-th SIB-RIS, (\ref{eq7d}) is the PS coefficient constraint, (\ref{eq7e}) represents the energy budget constraint of the SIB-RIS, and (\ref{eq7f}) guarantees the QoS of the PU.
	
	 The original problem (P1) exhibits intrinsic non-convexity and computational intractability due to two fundamental challenges: 1) The objective function incorporates a summation of logarithmic terms, introducing inherent non-convexity and analytical complexity; 2) The optimization variables are highly coupled within both the numerator and denominator of the objective function and its associated constraints, further exacerbating the computational difficulty in obtaining a globally optimal solution.
	
	\section{Proposed Joint Optimization Algorithm}
	 To address these challenges, we first reformulate the objective function into a more tractable form by employing the Lagrangian dual transformation and the quadratic transform techniques. Subsequently, a BCD-based joint optimization algorithm is developed to decompose the original problem (\rm P1) into four more manageable sub-problems.
	
	\subsection{Equivalent Transformation of Objective Function}
	Since $\log_2(x) = \ln(x)/\ln 2$, maximizing $\sum_{j=1}^M \omega_j \log_2(1+\gamma_j)$ is equivalent to maximizing $\sum_{j=1}^M \omega_j \ln(1+\gamma_j)$. Hence, without loss of optimality, we replace $\log_2(\cdot)$ with $\ln(\cdot)$ in the subsequent derivations.
	
	To handle both the logarithmic nonlinearity and the strong coupling among the optimization variables, we first apply the Lagrangian dual transform to the objective function of (\rm P1), which yields
		\begin{equation}
			\begin{aligned}
				&\sum_{j=1}^M \omega_j \ln(1 + \gamma_j)  \\
				= &\sum_{j=1}^M \omega_j 
				\max_{\alpha_j > 0}
				\left[
				\ln(1 + \alpha_j) - \alpha_j 
				+ \frac{(1 + \alpha_j)\gamma_j}{1 + \gamma_j}
				\right],
			\end{aligned}
			\label{eq8}
		\end{equation}
	where $\boldsymbol{\alpha} = [\alpha_1,\ldots,\alpha_M]^T$ is an auxiliary variable vector and the equality in \eqref{eq8} holds with $\alpha_j^\star = \gamma_j$. For any given $\bm{\alpha}$, the WSSE maximization objective is equivalent to
	\[
	\max_{{\bf{w}},{\cal R},{\bm \delta}} 
	\sum_{j = 1}^M 
	\frac{\omega_j\left(1 + \alpha_j\right)\gamma_j}{1 + \gamma_j}.
	\]
	
	This multiple-ratio fractional programming problem can be further recast, by means of the quadratic transform, into a biconvex optimization problem with the following equivalent objective function \cite{R35}:
	\begin{equation}
		\begin{aligned}
			f_1 & \!\left( {\bm \alpha} ,{\bm \beta} ,{\bf{w}},{\mathcal R},{\bm{\delta }} \right) 
			= \sum_{j = 1}^M \omega_j \big[ \ln(1 + \alpha_j) - \alpha_j \big] \\
			& \quad + \sum_{j = 1}^M 2\sqrt {\omega_j \left( 1 + \alpha_j \right)} 
			\,{\Re}\left\{ \beta_j^* A_j \right\}  
			- \sum_{j = 1}^M |\beta_j|^2 B_j,
		\end{aligned}
		\label{eq9_0}
	\end{equation}
	where $\bm{\beta} = [\beta_1,\ldots,\beta_M]^T$ is another auxiliary variable vector, and $A_j$ and $B_j$ are given by
	\begin{equation}
		\begin{aligned}
			{A_j} = {{\delta _j}}{\bf{g}}_j^H{\rm{diag}}\left( {{{\bf{F}}_j}{\bf{w}}} \right){{\bm \phi} _j},
		\end{aligned}
		\label{eq9}
	\end{equation}
	\begin{equation}
		\begin{aligned}
			{B_j} = \sum\limits_{i \ge j}^M {{{\left| {{{\delta _i}}{\bf{g}}_i^H{\rm{diag}}\left( {{{\bf{F}}_i}{\bf{w}}} \right){{\bm \phi} _i}} \right|}^2}}  + {\left| {{\bf{h}}^H{\bf{w}}} \right|^2} + {\sigma ^2}.
		\end{aligned}
		\label{eq10}
	\end{equation}
	
	Therefore, the initial WSSE optimization maximization problem can be equivalently rewritten as
	\begin{subequations}
		\begin{equation}
			\begin{aligned}
				\left( {{\rm{P2}}} \right)\mathop {\max }\limits_{{\bm{\alpha }},{\bm{\beta }},{\bm{w}},{\cal R},{\bm{\delta }}} {f_1}\left( {{\bm{\alpha }},{\bm{\beta }},{\bm{w}},{\cal R},{\bm{\delta }}} \right)
			\end{aligned}
			\label{eq11a}
		\end{equation}
		\begin{equation}
			\begin{aligned}
				{\mathrm{s.t.}} \ {\alpha _j} > 0,
			\end{aligned}
			\label{eq11b}
		\end{equation}
		\vspace{-6mm}
		\begin{equation}
			\begin{aligned}
				\qquad \ \ \eqref{eq7b}-\eqref{eq7f}.
			\end{aligned}
			\label{eq11c}
		\end{equation}
	\end{subequations}
	
	\subsection{BCD-based Joint Optimization Scheme}
	To address the interdependence among the optimization variables within both the objective function and the constraints, the BCD principle is employed. This approach adopts a cyclic update strategy, where the variables ${\bm \alpha}$, ${\bm \beta}$, ${\bf{w}}$, ${\mathcal R}$, and ${\bm \delta }$ are alternately optimized in each iteration. Under the BCD framework, problem (P2) can be decomposed into four sub-problems, which are sequentially solved as outlined in the following steps.
	
	1) \textit{Step 1 - Optimizing ${\bm \alpha}$ and ${\bm \beta}$ with Given $\left( {{\bf{w}},{\cal R},{\bm{\delta }}} \right)$:} For the given variables $\left( {{\bf{w}},{\cal R},{\bm{\delta }}} \right)$, the optimal $\left( {{{\bm{\alpha }}^*},{{\bm{\beta }}^*}} \right)$ can be acquired by setting the partial derivative of ${f_1}\left( {{\bm{\alpha }},{\bm{\beta }},{\bm{w}},{\cal R},{\bm{\delta }}} \right)$ with respect to ${\bm \alpha}$ and ${\bm \beta}$ equal to zero, which are expressed as
	\begin{equation}
		\begin{aligned}
			\alpha _j^* = {\gamma _j},\;\;\beta _j^* = \frac{{\sqrt {{\omega _j}\left( {1 + \alpha _j^*} \right)} {A_j}}}{{{B_j}}}.
		\end{aligned}
		\label{eq12}
	\end{equation}
	
%	Optimizing the coefficient vectors $\mathcal R$
	2) \textit{Step 2 - Reflection Coefficient Design for SIB-RIS}: When variables $\left( {{\bm{\alpha }},{\bm{\beta }},{\bm{w}},{\bm{\delta }}} \right)$ are given, the objective function can be reformulated as
	\begin{equation}
		\begin{aligned}
			& \qquad \qquad \qquad \qquad \quad \mathop {\max }\limits_{\mathcal R} {f_1} \Leftrightarrow   \\ 
			& \mathop {\max }\limits_{\cal R} \sum\limits_{j = 1}^M {\left[ {2\sqrt {{\omega _j}\left( {1 + {\alpha _j}} \right)} {{\Re}}\left\{ {\beta _j^*{{\bm \varepsilon} _j}{{\bm \phi} _j}} \right\} - {{\left| {{\beta _j}} \right|}^2}\sum\limits_{i \ge j}^M {{{\left| {{{\bm \varepsilon} _i}{{\bm \phi} _i}} \right|}^2}} } \right]},
		\end{aligned}
		\label{eq13}
	\end{equation} 
where $
		\boldsymbol{\varepsilon}_j 
		= \delta_j \mathbf{g}_j^H 
		\mathrm{diag}\!\left(\mathbf{F}_j \mathbf{w}\right)$ and $
		\boldsymbol{\varepsilon}_{j{\rm p}} 
		= \delta_j \mathbf{g}_{{\rm p}j}^H 
		\mathrm{diag}\!\left(\mathbf{F}_j \mathbf{w}\right)$ denote the effective cascaded channels from the PT to the AP and to the PU via the $j$-th SIB-RIS, respectively. For notational brevity, we introduce the lifted reflection vector and its auto-correlation matrix as $
		\hat{\boldsymbol{{\bm \phi}}}_j =
		\begin{bmatrix}
			\boldsymbol{{\bm \phi}}_j \\
			1
		\end{bmatrix},
		\hat{\boldsymbol{\Phi}}_j 
		= \hat{\boldsymbol{{\bm \phi}}}_j \hat{\boldsymbol{{\bm \phi}}}_j^H,$
		which satisfies $\hat{\boldsymbol{\Phi}}_j \succeq \mathbf{0}$ and 
		$\mathrm{rank}\!\left(\hat{\boldsymbol{\Phi}}_j\right) = 1$.
		
		Based on the above definitions, the linear term and the quadratic terms in \eqref{eq13} can be compactly expressed in quadratic matrix form by defining
		\[
		\boldsymbol{\Lambda}_{jj} =
		\begin{bmatrix}
			\mathbf{0}_{K\times K} 
			& \big(\sqrt{\omega_j(1+\alpha_j)}\,\beta_j^* \boldsymbol{\varepsilon}_j\big)^H \\
			\sqrt{\omega_j(1+\alpha_j)}\,\beta_j^* \boldsymbol{\varepsilon}_j 
			& 0
		\end{bmatrix},
		\]
		\[
		\boldsymbol{\Omega}_{jj} =
		\begin{bmatrix}
			\boldsymbol{\varepsilon}_j^H \boldsymbol{\varepsilon}_j & \mathbf{0}_{K\times 1} \\
			\mathbf{0}_{1\times K} & 0
		\end{bmatrix},
		\qquad
		\boldsymbol{\Omega}_{j{\rm p}} =
		\begin{bmatrix}
			\boldsymbol{\varepsilon}_{j{\rm p}}^H \boldsymbol{\varepsilon}_{j{\rm p}} & \mathbf{0}_{K\times 1} \\
			\mathbf{0}_{1\times K} & 0
		\end{bmatrix}.
		\]
		
		It then follows that
		\[
		2\sqrt{\omega_j(1+\alpha_j)}
		\,{\Re}\big\{\beta_j^* \boldsymbol{\varepsilon}_j \boldsymbol{{\bm \phi}}_j\big\}
		= \mathrm{Tr}\!\left(\boldsymbol{\Lambda}_{jj}\hat{\boldsymbol{\Phi}}_j\right),
		\]
		\[
		\big|\boldsymbol{\varepsilon}_j \boldsymbol{{\bm \phi}}_j\big|^2
		= \mathrm{Tr}\!\left(\boldsymbol{\Omega}_{jj}\hat{\boldsymbol{\Phi}}_j\right),
		\qquad
		\big|\boldsymbol{\varepsilon}_{j{\rm p}} \boldsymbol{{\bm \phi}}_j\big|^2
		= \mathrm{Tr}\!\left(\boldsymbol{\Omega}_{j{\rm p}}\hat{\boldsymbol{\Phi}}_j\right),
		\]
		which enables a unified semidefinite programming (SDP) formulation of the reflection-coefficient optimization problem. Then based on the quadratic constrained quadratic programming (QCQP) theory, problem (\rm P2) can be remodeled into
		\begin{subequations}
			\begin{align}
				\left(\mathrm{P3}\right)\ 
				\max_{\mathcal{R}}
				& \sum_{j=1}^{M}\Bigg[
				\mathrm{Tr}\!\left(\mathbf{\Lambda}_{jj}\,\hat{\mathbf{\Phi}}_j\right)
				- |\beta_j|^2 \sum_{i\ge j}^{M}
				\mathrm{Tr}\!\left(\mathbf{\Omega}_{ii}\,\hat{\mathbf{\Phi}}_i\right)
				\Bigg]
				\label{eq14a}\\
				\mathrm{s.t.}\quad
				& \hat{\mathbf{\Phi}}_j \succeq \mathbf{0},
				\label{eq14b}\\
				& \left[\hat{\mathbf{\Phi}}_j\right]_{k,k} = 1,\quad
				k \in \{1,\ldots,K+1\},
				\label{eq14c}\\
				& \mathrm{rank}\!\left(\hat{\mathbf{\Phi}}_j\right)=1,
				\label{eq14d}\\
				& \frac{\left|\mathbf{h}_{\mathrm{p}}^{H}\mathbf{w}\right|^{2}}
				{\sum_{j=1}^{M}\mathrm{Tr}\!\left(\mathbf{\Omega}_{j\mathrm{p}}\,\hat{\mathbf{\Phi}}_j\right)+\sigma^{2}}
				\ge 2^{R^{\mathrm{TH}}}-1.
				\label{eq14e}
			\end{align}
		\end{subequations}
    
    Notably, the remaining non-convexity arises from the rank-one constraint in (\ref{eq14d}). A widely adopted strategy is the semidefinite relaxation (SDR) method, which relaxes the rank-one constraint and reformulates the problem as a standard SDP problem. After obtaining the solution, a rank-one approximation is typically recovered via singular value decomposition (SVD) and Gaussian randomization. However, the SDR can be non-tight and yield a higher-rank optimal solution, in which case a rank-one feasible point is commonly constructed via eigen-decomposition followed by Gaussian randomization. Such randomization-based recovery generally incurs additional computational complexity and may result in a degraded objective value due to the nonzero relaxation gap, especially in large-scale settings. To overcome these fundamental limitations, we adopt a difference-of-convex (DC) programming framework that directly incorporates the rank constraint through strategic constraint reformulation, which enables a more effective handling of the rank-one constraint.
    
   Mathematically, the rank-one constraint in \eqref{eq14d} can be equivalently written as
    	\begin{equation}
    		\mathrm{Tr}\!\left(\hat{\boldsymbol{\Phi}}_j\right) - \sigma_{\max}\!\left(\hat{\boldsymbol{\Phi}}_j\right) \le 0,
    		\label{eq15}
    	\end{equation}
    	where $\sigma_{\max}(\mathbf{X})$ denotes the largest singular value of $\mathbf{X}$. Specifically, let $\lambda_i(\hat{\boldsymbol{\Phi}}_j)$ denote the $i$-th largest eigenvalue of $\hat{\boldsymbol{\Phi}}_j$. Since $\hat{\boldsymbol{\Phi}}_j \succeq \mathbf{0}$, we have
    	\[
    	\mathrm{Tr}(\hat{\boldsymbol{\Phi}}_j) - \sigma_{\max}(\hat{\boldsymbol{\Phi}}_j)
    	= \sum\nolimits_{i \ge 2} \lambda_i(\hat{\boldsymbol{\Phi}}_j) \ge 0,
    	\]
    	and the equality 
    	$\mathrm{Tr}(\hat{\boldsymbol{\Phi}}_j) - \sigma_{\max}(\hat{\boldsymbol{\Phi}}_j)=0$
    	holds if and only if $\mathrm{rank}(\hat{\boldsymbol{\Phi}}_j)=1$. Therefore, imposing
    	\eqref{eq15} together with $\hat{\boldsymbol{\Phi}}_j \succeq \mathbf{0}$ is equivalent to the rank-one constraint in \eqref{eq14d}. Moreover, 
    	$\sigma_{\max}(\hat{\boldsymbol{\Phi}}_j)=\|\hat{\boldsymbol{\Phi}}_j\|_2$
    	is a spectral function and is convex with respect to $\hat{\boldsymbol{\Phi}}_j$. As a result, the left-hand side of \eqref{eq15} is expressed as a linear function minus a convex function, which admits a convenient DC representation.
    	
    	Nevertheless, the reformulated constraint remains non-convex due to the non-smoothness of $\sigma_{\max}(\hat{\boldsymbol{\Phi}}_j)$. To tackle this challenge, we adopt a first-order Taylor expansion and successive convex approximation (SCA) to construct an affine global under-estimator:
    	\begin{equation}
    		\sigma_{\max}(\hat{\boldsymbol{\Phi}}_j) 
    		\ge \sigma_{\max}(\hat{\boldsymbol{\Phi}}_j^l) 
    		+ \mathrm{Tr}\!\big( \mathbf{v}_{\max}\mathbf{v}_{\max}^H 
    		\left( \hat{\boldsymbol{\Phi}}_j - \hat{\boldsymbol{\Phi}}_j^l \right) \big),
    		\label{eq16}
    	\end{equation}
    	where $\hat{\boldsymbol{\Phi}}_j^l$ denotes the feasible point at the $l$-th iteration and $\mathbf{v}_{\max}$ is the eigenvector associated with $\sigma_{\max}(\hat{\boldsymbol{\Phi}}_j^l)$. Consequently, the rank-one constraint is conservatively enforced by the following convex constraint:
    	\begin{equation}
    		\mathrm{Tr}\!\left(\hat{\boldsymbol{\Phi}}_j\right) 
    		- \sigma_{\max}\!\left(\hat{\boldsymbol{\Phi}}_j^l\right)
    		- \mathrm{Tr}\!\big( \mathbf{v}_{\max}\mathbf{v}_{\max}^H 
    		\left( \hat{\boldsymbol{\Phi}}_j - \hat{\boldsymbol{\Phi}}_j^l \right) \big)
    		\le 0.
    		\label{eq17}
    	\end{equation}	    	
    
    However, securing an initial feasible solution for ${{\bf{\hat \Phi }}}_j$
    poses a significant challenge, since the rank-one condition is implicitly
    enforced by the non-convex DC constraint
    ${\rm{Tr}}\!\left( {{{{\bf{\hat \Phi }}}_j}} \right) - 
    {\sigma _{\max }}\!\left( {{{{\bf{\hat \Phi }}}_j}} \right) \le 0$
    together with ${{\bf{\hat \Phi }}}_j \succeq \mathbf 0$. To circumvent this limitation, we propose a penalty-based method to reformulated problem (\rm P3) as
%   Although the residual optimization is convex, the initial feasible set of the problem is hard to obtain.   
    \begin{subequations}
    	\begin{equation}
    		\begin{aligned}
    				\left( {{\rm{P3}}.{\rm{1}}} \right)\; & \mathop {\max }\limits_{\cal R} \sum\limits_{j = 1}^M {\left[ {{\rm{Tr}}\left( {{{\bf{\Lambda }}_{jj}}{{{\bf{\hat \Phi }}}_j}} \right) - |{\beta _j}{|^2}\sum\limits_{i \ge j}^M {{\rm{Tr}}\left( {{{\bf{\Omega }}_{ii}}{{{\bf{\hat \Phi }}}_i}} \right)} } \right]}  \\
    				& \qquad \qquad - \rho \sum\limits_{j = 1}^M {{\eta _j}} 
    		\end{aligned}
    		\label{eq18a}
    	\end{equation}
    	\begin{equation}
    		\begin{aligned}
    			{\mathrm{s.t.}} \ {\eta _j} \ge 0,
    		\end{aligned}
    		\label{eq18b}
    	\end{equation}
    	\begin{equation}
    		\begin{aligned}
    			\eqref{eq14b}-\eqref{eq14c}, \eqref{eq14e},
    		\end{aligned}
    		\label{eq18c}
    	\end{equation}
    	\begin{equation}
    		\begin{aligned}
    			{\rm{Tr}}\left( {{{{\bf{\hat \Phi }}}_j}} \right) - {\sigma _{\max }}\left( {{\bf{\hat \Phi }}_j^l} \right) - {\rm{Tr}}\left( {{{\bf{v}}_{\max }}{{\bf{v}}_{{{\max }}}^H}\left( {{{{\bf{\hat \Phi }}}_j} - {\bf{\hat \Phi }}_j^l} \right)} \right) \le {\eta _j},
    		\end{aligned}
    		\label{eq18d}
    	\end{equation}
    \end{subequations}
    where $\rho > 0 $ denotes the penalty coefficient. The reconstructed optimization problem (\rm P3.1) is a standard SDP form, which can be efficiently solved using the CVX toolbox. The penalty term $\rho$ is initially set to a small value to prioritize feasibility attainment and is progressively increased to a sufficiently large magnitude. This adaptive strategy ensures two purposes: 1) it enforces strict satisfaction of constraint (\ref{eq18d}) by penalizing deviations from rank-one feasibility; and 2) it steers the optimization process toward a near-optimal solution. As $\rho  \to \infty $, the reformulated problem asymptotically enforces the original rank-one constraint, while maintaining computational tractability during initialization via controlled relaxation.
	
%	optimizing the beamforming vector $\bm w$
	3) \textit{Step 3 - Transmit Beamforming Design for PT }: Given $\left( \boldsymbol{\alpha}, \boldsymbol{\beta}, \mathcal{R}, \boldsymbol{\delta} \right)$, problem (\rm P2) reduces to the following optimization problem with respect to the transmit beamforming vector $\mathbf{w}$:
	\begin{subequations}
		\begin{align}
			\left(\mathrm{P4}\right)\ 
			\max_{\mathbf{w}}\ 
			& \sum_{j=1}^{M} 2\sqrt{\omega_j\left(1+\alpha_j\right)}
			\,\Re\!\left\{\beta_j^{*}\,\boldsymbol{\xi}_{j}\mathbf{w}\right\}
			\nonumber\\
			& \quad
			- \sum_{j=1}^{M} |\beta_j|^{2}
			\left[
			\sum_{i\ge j}^{M} \left|\boldsymbol{\xi}_{i}\mathbf{w}\right|^{2}
			+ \left|\mathbf{h}^{H}\mathbf{w}\right|^{2}
			\right]
			\label{eq19a}\\
			\mathrm{s.t.}\quad
			& \mathrm{Tr}\!\left(\mathbf{w}\mathbf{w}^{H}\right) \le P,
			\label{eq19b}\\
			& \chi\!\left(1-\delta_j^{2}\right)\left\|\mathbf{F}_j\mathbf{w}\right\|^{2}
			\ge \mu K,
			\label{eq19c}\\
			& \frac{\left|\mathbf{h}_{\mathrm{p}}^{H}\mathbf{w}\right|^{2}}
			{\sum_{j=1}^{M}\left|\boldsymbol{\xi}_{j\mathrm{p}}\mathbf{w}\right|^{2}+\sigma^{2}}
			\ge 2^{R^{\mathrm{TH}}}-1,
			\label{eq19d}
		\end{align}
	\end{subequations}
	 where $\boldsymbol{\xi}_j 
	 	= \delta_j \mathbf{g}_j^H \mathrm{diag}(\boldsymbol{{\bm \phi}}_j)\mathbf{F}_j$ and 
	 	$\boldsymbol{\xi}_{j{\rm p}} 
	 	= \delta_j \mathbf{g}_{{\rm p}j}^H \mathrm{diag}(\boldsymbol{{\bm \phi}}_j)\mathbf{F}_j$ denote the effective cascaded PT-to-SIB-RIS-to-AP and PT-to-SIB-RIS-to-PU channels associated with the $j$-th SIB-RIS, respectively. For notational simplicity, we introduce the lifted beamforming vector $\hat{\mathbf{w}} = [\mathbf{w}^T,\,1]^T$ and its covariance matrix 
	 	$\hat{\mathbf{W}} = \hat{\mathbf{w}}\hat{\mathbf{w}}^H \succeq \mathbf{0}$, and the direct PT-AP and PT-PU channels are represented by the block matrices
	 	\begin{equation}
	 		\mathbf{H} =
	 		\begin{bmatrix}
	 			\mathbf{h}\mathbf{h}^H & \mathbf{0}_{N\times 1} \\
	 			\mathbf{0}_{1\times N} & 0
	 		\end{bmatrix},
	 		\qquad
	 		\mathbf{H}_{\rm p} =
	 		\begin{bmatrix}
	 			\mathbf{h}_{\rm p}\mathbf{h}_{\rm p}^H & \mathbf{0}_{N\times 1} \\
	 			\mathbf{0}_{1\times N} & 0
	 		\end{bmatrix}, \nonumber
	 	\end{equation}
	 	so that
	 	\begin{equation}
	 		|\mathbf{h}^H\mathbf{w}|^2 = \mathrm{Tr}(\mathbf{H}\hat{\mathbf{W}}), 
	 		\qquad
	 		|\mathbf{h}_{\rm p}^H\mathbf{w}|^2 = \mathrm{Tr}(\mathbf{H}_{\rm p}\hat{\mathbf{W}}). \nonumber
	 	\end{equation}
	 	
	 	Furthermore, we define the auxiliary matrices
	 		 	\[
	 	\mathbf{N}_{jj}=
	 	\begin{bmatrix}
	 		\mathbf{0}_{N\times N} & \big(\beta_j^{*}\boldsymbol{\xi}_j\big)^{H}\\
	 		\beta_j^{*}\boldsymbol{\xi}_j & 0
	 	\end{bmatrix},
	 	\qquad
	 	\mathbf{T}_j=
	 	\begin{bmatrix}
	 		\mathbf{F}_j^{H}\mathbf{F}_j & \mathbf{0}\\
	 		\mathbf{0} & 0
	 	\end{bmatrix},
	 	\]
	 	\begin{equation}
	 		\mathbf{M}_{jj} = 
	 		\begin{bmatrix}
	 			\boldsymbol{\xi}_j^H \boldsymbol{\xi}_j & \mathbf{0}_{N\times 1} \\
	 			\mathbf{0}_{1\times N} & 0
	 		\end{bmatrix},
	 		\qquad
	 		\mathbf{M}_{j{\rm p}} = 
	 		\begin{bmatrix}
	 			\boldsymbol{\xi}_{j{\rm p}}^H \boldsymbol{\xi}_{j{\rm p}} & \mathbf{0}_{N\times 1} \\
	 			\mathbf{0}_{1\times N} & 0
	 		\end{bmatrix}, \nonumber
	 	\end{equation}
	 	which yields 
	 	\begin{equation}
	 		2{\Re}\{\beta_j^* \boldsymbol{\xi}_j\mathbf{w}\}
	 		= \mathrm{Tr}(\mathbf{N}_{jj} \hat{\mathbf{W}}),
	 	\end{equation}
	 	\begin{equation}
	 		|\boldsymbol{\xi}_j\mathbf{w}|^2 = \mathrm{Tr}(\mathbf{M}_{jj}\hat{\mathbf{W}}),
	 		\quad
	 		|\boldsymbol{\xi}_{j{\rm p}}\mathbf{w}|^2 = \mathrm{Tr}(\mathbf{M}_{j{\rm p}}\hat{\mathbf{W}}),
	 	\end{equation}
	 	and
	 	\begin{equation}
	 		\|\mathbf{F}_j \mathbf{w}\|^2 = \mathrm{Tr}(\mathbf{T}_j \hat{\mathbf{W}}).
	 	\end{equation}
	 
	 Then, based on the QCQP and penalty-based techniques, problem (\rm P4) can be equivalently transformed into 
\begin{subequations}
	
	\begin{align}
		\left(\mathrm{P4.1}\right)\ 
		\max_{\hat{\mathbf{W}}}\ 
		& \sum_{j=1}^{M}\sqrt{\omega_j\left(1+\alpha_j\right)}
		\,\mathrm{Tr}\!\left(\mathbf{N}_{jj}\hat{\mathbf{W}}\right)
		- \rho_{1}\upsilon
		\nonumber\\
		& 
		- \sum_{j=1}^{M}|\beta_j|^{2}
		\left[
		\sum_{i\ge j}^{M}\mathrm{Tr}\!\left(\mathbf{M}_{ii}\hat{\mathbf{W}}\right)
		+ \mathrm{Tr}\!\left(\mathbf{H}\hat{\mathbf{W}}\right)
		\right],
		\label{eq20a}\\
		\mathrm{s.t.}\quad
		& \mathrm{Tr}\!\left(\mathbf{\Xi}\hat{\mathbf{W}}\right) \le P,
		\label{eq20b}\\
		& \chi\!\left(1-\delta_j^{2}\right)\mathrm{Tr}\!\left(\mathbf{T}_j\hat{\mathbf{W}}\right)
		\ge \mu K,
		\label{eq20c}\\
		& \frac{\mathrm{Tr}\!\left(\mathbf{H}_{\mathrm{p}}\hat{\mathbf{W}}\right)}
		{\sum_{j=1}^{M}\mathrm{Tr}\!\left(\mathbf{M}_{j\mathrm{p}}\hat{\mathbf{W}}\right)+\sigma^{2}}
		\ge 2^{R^{\mathrm{TH}}}-1,
		\label{eq20d}\\
		& \hat{\mathbf{W}} \succeq \mathbf{0},\quad
		\left(\hat{\mathbf{W}}\right)_{N+1,N+1}=1,
		\label{eq20f}\\
		& \upsilon \ge 0,
		\label{eq20h}
	\end{align}
	
	\begin{equation}
		\mathrm{Tr}\!\left(\hat{\mathbf{W}}\right)
		- \sigma_{\max}\!\left(\hat{\mathbf{W}}^{\,l}\right)
		- \mathrm{Tr}\!\left(
		\mathbf{u}_{\max}\mathbf{u}_{\max}^{H}
		\left(\hat{\mathbf{W}}-\hat{\mathbf{W}}^{\,l}\right)
		\right)
		\le \upsilon,
		\label{eq20g}
	\end{equation}
\end{subequations}
	 where ${{\rho _1}}>0$ is the introduced penalty factor and 
	 \[{\bf{\Xi }} = \left[ {\begin{array}{*{20}{l}}
	 		{{\bf{I}},{\rm{   }}{\bf{0}}}\\
	 		{{\bf{0}},\;\;\;0}
	 \end{array}} \right],{{\bm{{\rm T}}}_j} = \left[ {\begin{array}{*{20}{l}}
	 		{{\bf{F}}_j^H{{\bf{F}}_j},{\bf{0}}}\\
	 		{\quad {\bf{0}},\;\;\;0}
	 \end{array}} \right].\]
 
     The optimal solution for the problem (\rm P4.1) can be efficiently acquired by CVX toolbox since it is a standard SDP.  
   
%	optimizing the PS vector $\bm \delta$ 
	4) \textit{Step 4 - PS Vector Design}: For given $\left( {{\bm{\alpha }},{\bm{\beta }},{\bm w},{\mathcal{R}}} \right)$,  the optimization problem (\rm P2) can be equivalently converted into
	\begin{subequations}
		\begin{equation}
			\begin{aligned}
				\left( {{\rm{P5}}} \right) \ \mathop {\max }\limits_{\bm{\delta }} \sum\limits_{j = 1}^M {{\delta _j}2\sqrt {{\omega _j}\left( {1 + {\alpha _j}} \right)} {{\Re}}\left\{ {\beta _j^*{\bf{g}}_j^H{\rm{diag}}\left( {{{\bm \phi} _j}} \right){{\bf{F}}_j}{\bf{w}}} \right\}} \\
				- \sum\limits_{j = 1}^M {{\left| {{\beta _j}} \right|^2}\left[ {\sum\limits_{i \ge j}^M {\delta _i^2{{\left| {{\bf{g}}_i^H{\rm{diag}}\left( {{{\bm \phi} _i}} \right){{\bf{F}}_i}{\bf{w}}} \right|}^2}} } \right]} 
			\end{aligned}
			\label{eq21a}
		\end{equation}
		\begin{equation}
			\begin{aligned}
				{\mathrm{s.t.}} \ \eqref{eq7d}-\eqref{eq7f}.
			\end{aligned}
			\label{eq21b}
		\end{equation}
	\end{subequations}
	
%	{\color{blue}\textit{Remark 1 (Convexity of (P5)):} 
%		For fixed $\left(\boldsymbol{\alpha}, \boldsymbol{\beta}, \mathbf{w}, \mathcal{R}\right)$, 
%		problem (P5) is convex with respect to the PS vector $\boldsymbol{\delta}$. 
%		Specifically, the objective in \eqref{eq21a} can be rewritten as
%		$f(\boldsymbol{\delta}) = \sum_{j=1}^M a_j \delta_j - \sum_{i=1}^M c_i \delta_i^2$ 
%		with $c_i \ge 0$, and is thus concave in $\boldsymbol{\delta}$. 
%		The box constraints in \eqref{eq7d} and the EH constraints in \eqref{eq7e} define
%		convex intervals for each $\delta_j$. 
%		Moreover, the PU QoS constraint in \eqref{eq7f} is equivalent to a convex quadratic
%		inequality of the form $\sum_{i=1}^M k_i \delta_i^2 \le C$, with $k_i \ge 0$ and
%		$C$ being a constant determined by $\mathbf{w}$. 
%		Hence, (P5) maximizes a concave function over a convex feasible set and is therefore
%		a convex program, which can be efficiently solved by standard solvers such as CVX~\cite{Boyd2004}.
%		
%	}
	\textit{Lemma 1}: For any given $\left( \boldsymbol{\alpha}, \boldsymbol{\beta}, \mathbf{w}, \mathcal{R} \right)$, problem (P5) is a convex optimization problem (i.e., it maximizes a concave objective over a convex feasible set) and can therefore be efficiently solved by the CVX toolbox.
	
	\emph{Proof}: See Appendix~A.

\subsection{Computational Complexity Analysis}
To solve problem (P1), we develop an joint optimization algorithm that iteratively optimizes the $\left(\bm\alpha,\bm\beta,\mathcal R,\mathbf w,\bm\delta\right)$ through the four steps, the detailed computation procedure is given in \textbf{Algorithm \ref{alg:BCD}}. In Step~1, the optimal auxiliary variables $\bm\alpha$ and $\bm\beta$ are obtained from the closed-form expressions in \eqref{eq12}. Accordingly, the computational complexities of acquiring $\bm\alpha$ and $\bm\beta$ can be approximated as ${\mathcal O}\!\left(MKN\right)$ and ${\mathcal O}\!\left(MKN\right)$, respectively. In Step~2, the SIB-RIS reflection coefficients $\mathcal R$ are updated by solving subproblem (P3.1), which is a standard SDP and can be efficiently handled by an interior-point method with complexity ${\mathcal O}\!\left(M (K+1)^{4.5}\right)$ per DC iteration \cite{Complexity}. In Step~3, the active beamforming vector at the PT is obtained by solving subproblem (P4.1), whose standard SDP form can similarly be solved via an interior-point method with complexity ${\mathcal O}\!\left((N+1)^{4.5}\right)$ per DC iteration \cite{Complexity}. In Step~4, the PS vector $\bm\delta$ is optimized by solving (P5), which is a convex quadratic program, and its complexity is on the order of ${\mathcal O}\!\left(M^{3}\right)$ and is generally dominated by the SDP subproblems. Then, the overall computational complexity of the proposed scheme can be approximated as
\begin{equation}
	\mathcal O\!\left(
	I_1\left(
	2MKN
	+ I_2 M (K+1)^{4.5}
	+ I_3 (N+1)^{4.5}
	+ M^3
	\right)
	\right),
\end{equation}
where $I_1$, $I_2$, and $I_3$ denote the required numbers of iterations for convergence of the outer BCD loop, the inner DC procedure for (P3.1), and the inner DC procedure for (P4.1), respectively.

\begin{algorithm}[t]
	\caption{Proposed Joint Optimization Algorithm}
	\label{alg:BCD}
	\begin{algorithmic}[1]
		\STATE \textbf{Initialization:} Choose feasible $\mathbf{w}^{(0)}$ and $\mathcal{R}^{(0)}$; initialize penalty parameters $\rho^{(0)}$, $\rho_{1}^{(0)}$, scaling factor $c>1$, and maximum penalty $\rho_{\max}$. Set outer iteration index $t=0$ and compute $R_{\rm sum}^{(0)}$ via~\eqref{eq7a}.
		
		\STATE \textbf{Repeat} \hfill (outer BCD loop)
		% -------------------- Step 1 --------------------
		\STATE \hspace*{0.03in}\textbf{Step 1}: Update ${\boldsymbol{\alpha}}^{(t+1)}$ and ${\boldsymbol{\beta}}^{(t+1)}$ according to~\eqref{eq12}.
		
		% -------------------- Step 2 --------------------
		\STATE \hspace*{0.03in}\textbf{Step 2 (Reflection coefficient design for SIB-RIS):}
		\STATE \hspace*{0.12in}\textbf{Initialization (inner DC loop for $\mathcal{R}$):} Set $\rho \leftarrow \rho^{(0)}$ \hspace*{0.07in} and inner iteration index $l_1=0$.
		\STATE \hspace*{0.12in}\textbf{Repeat}
		\STATE \hspace*{0.18in}Update $\mathcal{R}^{(t,l_1+1)}$ and $\boldsymbol{\eta}^{(l_1+1)}$ by solving (P3.1).
		\STATE \hspace*{0.18in}Update $\rho \leftarrow \min\!\left\{ c\,\rho, \rho_{\max} \right\}$ and set $l_1 \leftarrow l_1+1$.
		\STATE \hspace*{0.12in}\textbf{Until} (P3.1) converges.
		\STATE \hspace*{0.12in}Set $\mathcal{R}^{(t+1)} \leftarrow \mathcal{R}^{(t,l_1)}$.
		
		% -------------------- Step 3 --------------------
		\STATE \hspace*{0.03in}\textbf{Step 3 (Transmit beamforming design for PT):}
		\STATE \hspace*{0.12in}\textbf{Initialization (inner DC loop for $\mathbf{w}$):} Set $\rho_1 \leftarrow \rho_1^{(0)}$ \hspace*{0.07in} and inner iteration index $l_2=0$.
		\STATE \hspace*{0.12in}\textbf{Repeat}
		\STATE \hspace*{0.18in}Update $\mathbf{w}^{(t,l_2+1)}$ and $\upsilon^{(l_2+1)}$ by solving (P4.1).
		\STATE \hspace*{0.18in}Update $\rho_1 \leftarrow \min\!\left\{ c\,\rho_1, \rho_{\max} \right\}$ and set $l_2 \leftarrow l_2+1$.
		\STATE \hspace*{0.12in}\textbf{Until} (P4.1) converges.
		\STATE \hspace*{0.12in}Set $\mathbf{w}^{(t+1)} \leftarrow \mathbf{w}^{(t,l_2)}$.
		
		% -------------------- Step 4 --------------------
		\STATE \hspace*{0.03in}\textbf{Step 4 (PS vector update):} Update $\boldsymbol{\delta}^{(t+1)}$ by solving (P5).
		
		\STATE \hspace*{0.03in}Compute $R_{\rm sum}^{(t+1)}$ via~\eqref{eq7a} and set $t \leftarrow t+1$.
		\STATE \textbf{Until} $\displaystyle\frac{R_{\rm sum}^{(t)} - R_{\rm sum}^{(t-1)}}{R_{\rm sum}^{(t-1)}} \le 0.01$.
		\STATE \textbf{Return} optimized $\mathcal{R}^\star$, $\mathbf{w}^\star$, and $\boldsymbol{\delta}^\star$.
	\end{algorithmic}
\end{algorithm}

\section{Simulation Results}
	In this section, simulation results are provided to validate the effectiveness of the SIB-RIS-based NOMA CR system and to demonstrate the superiority of the presented joint optimization scheme. 
\subsection{Simulation Setup}
In our simulations, we consider a three-dimensional (3D) coordinate system where the PT is located at $(0, 0, 10)$~m and $M=4$ SIB-RISs are uniformly distributed within a cylindrical region centered at $(5, 0)$~m with a radius of $10$~m, where the height of each SIB-RIS is randomly generated between $7$~m and $10$~m. The PT is assumed to be equipped with a uniform linear array (ULA) along the $x$-axis, while the SIB-RISs are assumed to deploy a uniform planar array (UPA) in $xz$-plane with $K_x=5$ columns and $K_z$ rows (${K_z}{\rm{ = }}\frac{K}{{{K_x}}}$). The AP is fixed at $(20, 15, 1)$~m, while the PU is uniformly distributed in a circular ground area centered at the origin with a radius of $20$~m.

The large-scale path loss (in dB) is modeled as
\begin{equation}
	\beta=\beta_0-10\alpha\log_{10}\!\left(\frac{d}{d_0}\right),
\end{equation}
where $\beta_0=-20$ dB is the path-loss at the reference distance $d_0=1$~m, $\alpha$ is the path-loss exponent, and $d$ is the link distance. To account for severe signal attenuation caused by obstacles, the path loss exponents for PT-to-AP and SIB-RIS-to-PU links are set to $\alpha_{\mathrm{PT,AP}}=3.5$ and $\alpha_{\mathrm{SIBRIS,PU}}=2.8$. In contrast, assuming that the SIB-RISs can be strategically deployed to ensure line-of-sight (LoS) connectivity, the exponents for the assistance links are set to $\alpha_{\mathrm{PT,SIBRIS}}=2.2$, $\alpha_{\mathrm{SIBRIS,AP}}=2.8$, and $\alpha_{\mathrm{PT,PU}}=2.8$, respectively.

Regarding the small-scale fading, we assume that all communication channels undergo Rician fading characterized by a Rician factor of $\kappa = 3$. The small-scale fading vector for the direct link between the PT and the AP is modeled as:
\begin{equation}
	\mathbf{\tilde h} = \sqrt{\frac{\kappa}{\kappa + 1}} \mathbf{\tilde h}^{\mathrm{LoS}} + \sqrt{\frac{1}{\kappa + 1}} \mathbf{\tilde h}^{\mathrm{NLoS}},
\end{equation}
where $\mathbf{\tilde h}^{\mathrm{NLoS}} \sim \mathcal{CN}(\mathbf{0}, \mathbf{I}_N)$ represents the non-line-of-sight (NLoS) Rayleigh fading component. The LoS component $\mathbf{\tilde h}^{\mathrm{LoS}}$ is determined by the transmit array geometry and is expressed as $\mathbf{\tilde h}^{\mathrm{LoS}} = \mathbf{a}_{\mathrm{PT}}(\phi^{\mathrm{AoD}}_{\mathrm{PT,AP}})$, where $\mathbf{a}_{\mathrm{PT}}(\phi)$ denotes the array response vector of the ULA at the PT, which is defined as
\begin{equation}
	\mathbf{a}_{\mathrm{PT}}(\phi) = \left[ 1, e^{-j \frac{2\pi \Delta}{\lambda} \cos\phi}, \dots, e^{-j \frac{2\pi \Delta}{\lambda} (N-1) \cos\phi} \right]^T,
	\label{eq:ULA_response}
\end{equation}
where $\lambda$ and $\Delta$ denote the carrier wavelength and antenna spacing, respectively, while $\phi^{\mathrm{AoD}}_{\mathrm{PT,AP}}$ represents the angle of departure (AoD). Similarly, the direct channel from the PT to the PU follows an identical distribution and is omitted for brevity.

\begin{figure*}[htpb]
	\centering
	\begin{minipage}[htbp]{0.46\linewidth}
		\centering
		\includegraphics[width=\linewidth]{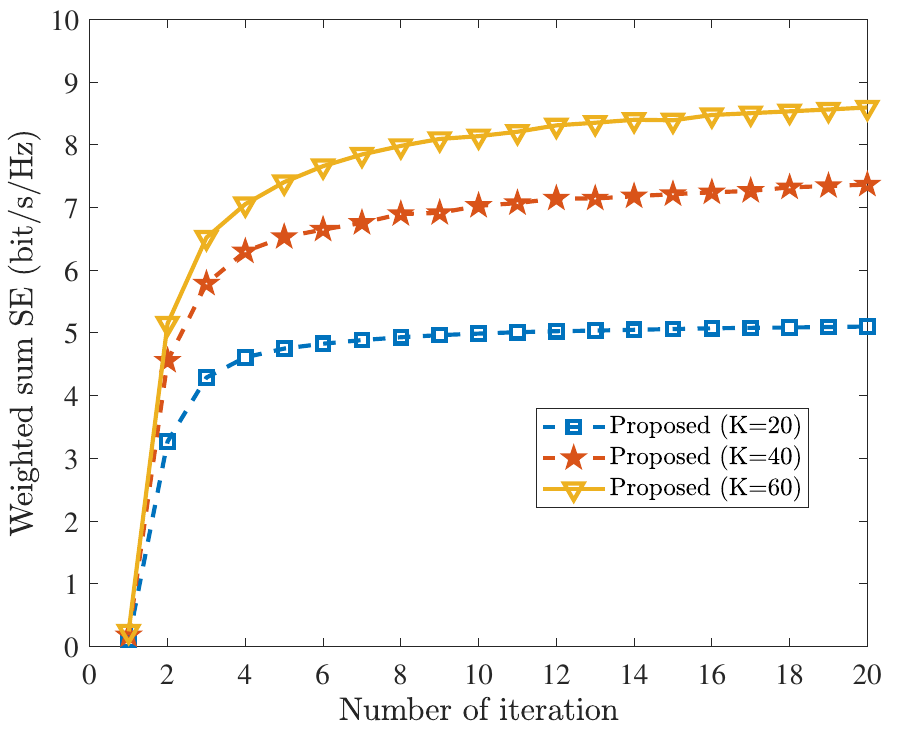}
		\caption{WSSE versus the number of iterations.}
		\label{F2}
	\end{minipage}
	\hfill
	\begin{minipage}[htbp]{0.46\linewidth}
		\centering
		\includegraphics[width=\linewidth]{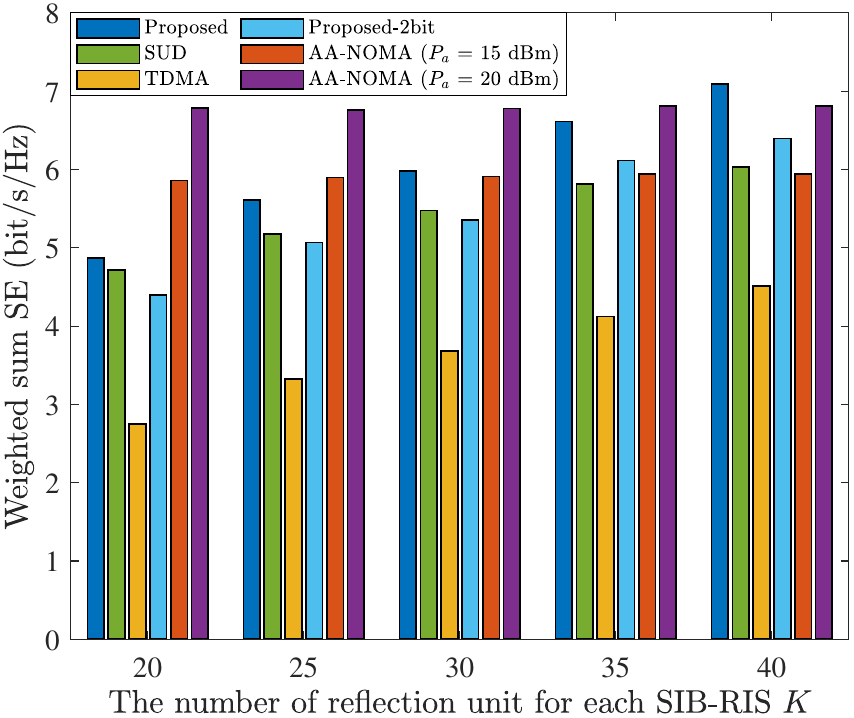}
		\caption{WSSE versus the number of reflection elements for each SIB-RIS.}
		\label{F3}
	\end{minipage}
\end{figure*}

The small-scale fading matrix corresponding to the link from the PT to the $j$-th SIB-RIS is given by
\begin{equation}
	\mathbf{\tilde F}_j = \sqrt{\frac{\kappa}{\kappa + 1}} \mathbf{\tilde F}_j^{\mathrm{LoS}} + \sqrt{\frac{1}{\kappa + 1}} \mathbf{\tilde F}_j^{\mathrm{NLoS}},
\end{equation}
where the LoS component $\tilde{\mathbf{F}}_j^{\mathrm{LoS}}$ depends on the azimuth AoD at the PT, denoted by $\phi_{\mathrm{PT},j}^{\mathrm{AoD}}$, and the azimuth and elevation angles of arrival (AoAs) at the SIB-RIS, denoted by $\phi_{j}^{\mathrm{AoA}}$ and $\theta_{j}^{\mathrm{AoA}}$, respectively. It is formulated as $\tilde{\mathbf{F}}_j^{\mathrm{LoS}} = \mathbf{a}_{\mathrm{RIS}}(\phi_{j}^{\mathrm{AoA}}, \theta_{j}^{\mathrm{AoA}}) \mathbf{a}_{\mathrm{PT}}^{H}(\phi_{\mathrm{PT},j}^{\mathrm{AoD}})$. Due to the SIB-RIS employs a UPA, its array response vector can be defined as $\mathbf{a}_{\mathrm{RIS}}(\phi, \theta) = \mathbf{a}_x(\phi, \theta) \otimes \mathbf{a}_z(\theta)$, where $\otimes$ denotes the Kronecker product and $\phi$ and $\theta$ are the azimuth and elevation angles of arrival/departure, respectively. The steering vectors along the $x$- and $z$-axes are respectively given by
\begin{align}
	\mathbf{a}_x(\phi, \theta) &= \left[1, e^{-j \frac{2\pi \Delta}{\lambda} \sin\theta \cos\phi}, \dots, e^{-j \frac{2\pi \Delta}{\lambda} (K_x-1) \sin\theta \cos\phi}\right]^T, \\
	\mathbf{a}_z(\theta)       &= \left[1, e^{-j \frac{2\pi \Delta}{\lambda} \cos\theta},           \dots, e^{-j \frac{2\pi \Delta}{\lambda} (K_z-1) \cos\theta}\right]^T.
\end{align}
where $\phi$ and $\theta$ represent the azimuth and elevation angles, respectively.

Furthermore, the small-scale fading vector from the $j$-th SIB-RIS to the AP is modeled as:
\begin{equation}
	\mathbf{g}_{j} = \sqrt{\frac{\kappa}{\kappa + 1}} \mathbf{g}_{j}^{\mathrm{LoS}} + \sqrt{\frac{1}{\kappa + 1}} \mathbf{g}_{j}^{\mathrm{NLoS}},
\end{equation}
where the LoS component is determined by the azimuth and elevation AoDs at the SIB-RIS towards the AP, denoted by $\phi_{j}^{\mathrm{AoD}}$ and $\theta_{j}^{\mathrm{AoD}}$, respectively. Consequently, the LoS component is expressed as $\mathbf{g}_{j}^{\mathrm{LoS}} = \mathbf{a}_{\mathrm{RIS}}(\phi_{j}^{\mathrm{AoD}}, \theta_{j}^{\mathrm{AoD}})$. Analogously, the channel from the SIB-RIS to the PU follows an identical Rician fading structure, and its detailed description is omitted for brevity.

Unless otherwise mentioned, the simulation parameters are set as follows: The PT is equipped with $N=4$ antennas to serve multiple SIB-RISs. The transmit power budget at the PT is set to $34~\mathrm{dBm}$, while the noise power is $\sigma^2 = -80~\mathrm{dBW}$. The energy harvesting efficiency is $\chi = 0.8$ with a power consumption constant of $\mu = 10^{-6}~\mathrm{W}$. Furthermore, the QoS threshold for the PU is fixed at $1.5~\mathrm{bit/s/Hz}$.

For performance comparison, the following comparative schemes are considered:
\begin{itemize}
	\item \textbf{Proposed}: The joint optimization algorithm developed in Section~III.
	
    \item \textbf{SUD}: The single-antenna AP adopts a low-complexity single-user decoding (SUD) receiver, which decodes each $x_j$ by treating the signals of other users and the primary-link component as noise~\cite{Xu}.

	\item \textbf{TDMA}: This scheme represents the joint optimization scheme for the SIB-RIS-assisted time-division multiple access (TDMA) CR system, in which a frame of duration $T$ is partitioned into $M$ equal orthogonal time slots, each assigned to one SIB-RIS.
	
	\item \textbf{Proposed-2bit}: A discrete-phase counterpart of the \textbf{Proposed} scheme, in which the same joint optimization framework is applied but each SIB-RIS element is equipped with a 2-bit phase shifter, restricting the reflection phase to four discrete levels.
	
	\item \textbf{AA-NOMA}: A conventional active-antenna NOMA benchmark without SIB-RISs, where each ST is equipped with $4$ active transmit antennas and the maximum transmit power is set to either $P_a = 15$~dBm or $P_a = 20$~dBm, which are labelled as ``AA-NOMA $P_a=15$ dBm'' and ``AA-NOMA $P_a=20$ dBm''.
\end{itemize}

\subsection{Results and Analyses}
	\label{subsec:results}
	
	Fig.~\ref{F2} illustrates the convergence behavior of the \textbf{Proposed} algorithm. For all considered numbers of reflecting elements, the WSSE monotonically increases with the iteration index and converges to a steady value within approximately $10$-$12$ iterations. In addition, increasing the number of SIB-RIS reflecting elements yields a noticeable gain in the converged WSSE, while the required number of iterations is only marginally affected. This insensitivity of the convergence speed to the SIB-RIS dimension indicates that the \textbf{Proposed} algorithm exhibits favorable scalability and can be efficiently applied to large-scale RIS deployments.
	
	Fig.~\ref{F3} depicts the WSSE versus the number of reflecting elements $K$ per SIB-RIS for all considered schemes. It can be seen that the WSSE of all SIB-RIS-based designs increases with $K$, since a larger number of reflecting elements provides more spatial DoFs and enhances the passive beamforming gain. In particular, when $K=30$ and $K=40$, the proposed SIB-RIS-assisted NOMA architecture outperforms the conventional AA-NOMA benchmarks with $P_a=15$~dBm and $P_a=20$~dBm, while relying only on nearly passive components and thus avoiding the hardware cost and energy consumption associated with multiple active RF chains and batteries at the STs. Moreover, the \textbf{Proposed} scheme consistently achieves higher WSSE than the \textbf{SUD} and \textbf{TDMA} benchmarks, which confirms the performance advantage of the NOMA-based design.
	
	\begin{figure*}[htpb]
		\centering
		\begin{minipage}[htbp]{0.46\linewidth}
			\centering
			\includegraphics[width=\linewidth]{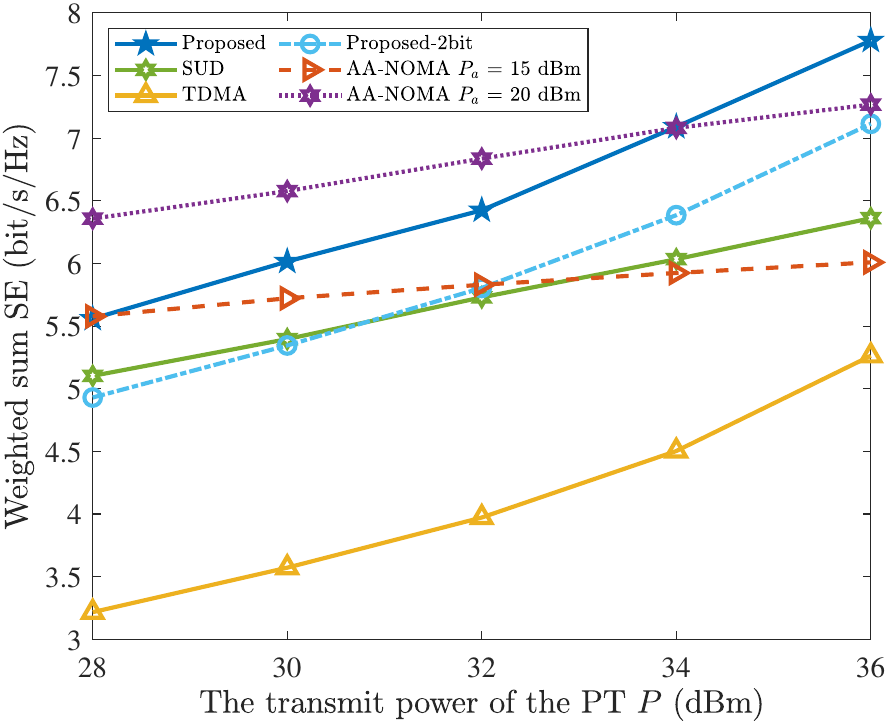}
			\caption{WSSE versus the transmit power of the PT.}
			\label{F4}
		\end{minipage}
		\hfill
		\begin{minipage}[htbp]{0.46\linewidth}
			\centering
			\includegraphics[width=\linewidth]{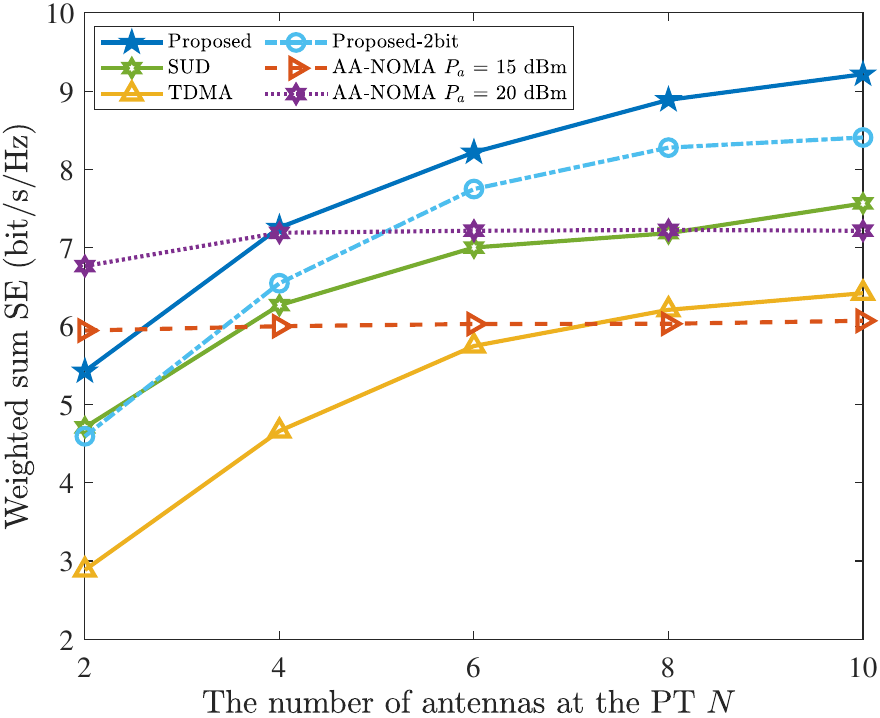}
			\caption{WSSE versus the number of antennas at the PT.}
			\label{F5}
		\end{minipage}
	\end{figure*}
	
	\begin{figure*}[!htpb]
		\centering
		\begin{minipage}[htbp]{0.46\linewidth}
			\centering
			\includegraphics[width=\linewidth]{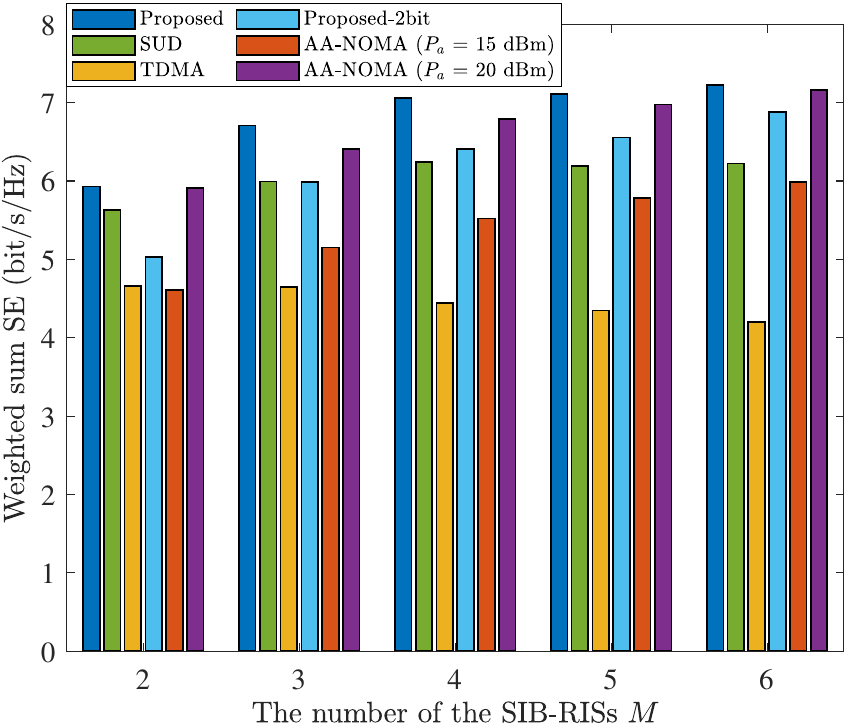}
			\caption{WSSE versus the number of SIB-RISs.}
			\label{F6}
		\end{minipage}
		\hfill
		\begin{minipage}[htbp]{0.46\linewidth}
			\centering
			\includegraphics[width=\linewidth]{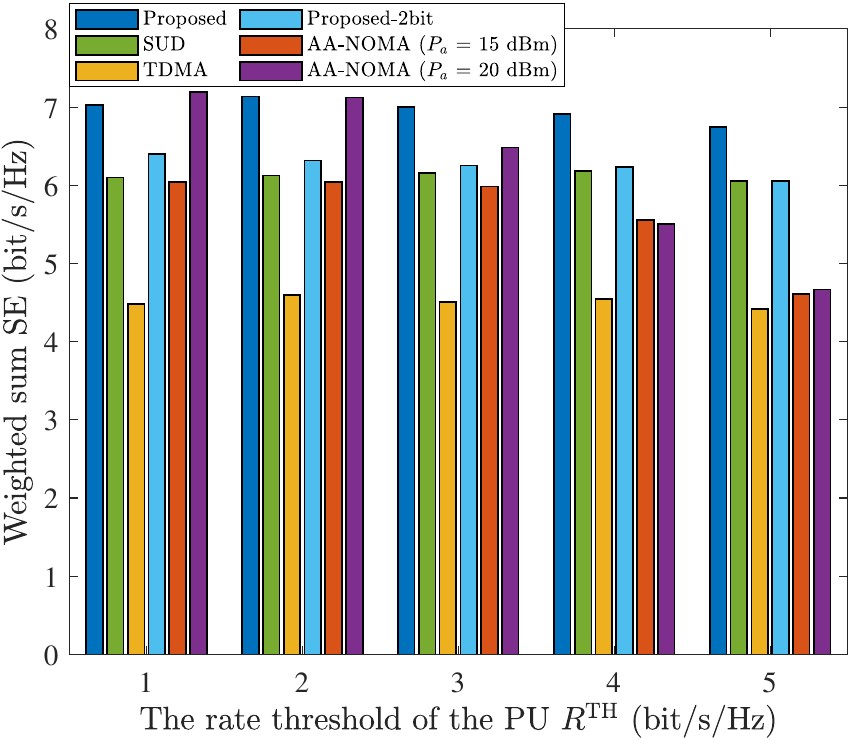}
			\caption{WSSE versus the rate threshold of the PU.}
			\label{F7}
		\end{minipage}
	\end{figure*}
	
	Fig.~\ref{F4} depicts the WSSE versus the PT transmit power budget $P$. For all schemes, the WSSE increases monotonically with $P$, since a higher transmit power enhances the incident signal power at each SIB-RIS, thereby enabling more harvested energy, relaxing the EH constraints, and allowing a larger fraction of the power to be used for information-bearing backscatter. The \textbf{Proposed} scheme yields the superior WSSE among all SIB-RIS-based designs and, in the moderate-to-high power regime (e.g., $P \ge 34$~dBm), also outperforms the AA-NOMA benchmark with $P_a=20$~dBm. It is further observed that, when $P \ge 34$~dBm, the \textbf{Proposed-2bit} scheme, despite the phase quantization loss, outperforms both the \textbf{SUD} benchmark and the \textbf{AA-NOMA} scheme with $P_a=15$~dBm. These results suggest that the proposed SIB-RIS architecture remains effective when implemented with practical low-resolution phase-shifter hardware, thereby supporting its engineering feasibility.
	
	Fig.~\ref{F5} investigates the impact of the number of PT antennas $N$ on the WSSE. As $N$ increases, all SIB-RIS-based schemes exhibit a clear performance improvement, since a larger transmit array provides additional spatial DoFs, enabling the beamformer $\mathbf{w}$ to more effectively concentrate energy toward the SIB-RISs while simultaneously strengthening the PT-PU link and thereby facilitating the satisfaction of the PU's QoS constraint. The \textbf{Proposed} scheme attains the highest WSSE over the entire range of $N$ and, when $N>4$, also outperforms the AA-NOMA benchmark with $P_a=20$~dBm. Furthermore, the \textbf{Proposed-2bit} scheme surpasses the AA-NOMA scheme with $P_a=20$~dBm for $N>6$, which demonstrates that a practical 2-bit SIB-RIS implementation can already provide competitive performance. 
	
	Fig.~\ref{F6} illustrates the WSSE versus the number of SIB-RISs. It is observed that the \textbf{Proposed} scheme consistently achieves the highest WSSE for all values of $M$, which confirms the effectiveness of the proposed design. In contrast, the WSSE performance of the \textbf{TDMA} scheme decreases monotonically with $M$, since the frame is equally partitioned into $M$ orthogonal time slots and the per-slot rate cannot compensate for the $1/M$ loss in time resources under a fixed PT power budget and PU QoS constraint, which is consistent with the inherent spectral-efficiency limitation of OMA.
	
	Fig.~\ref{F7} shows the WSSE of the SN as a function of the PU rate threshold $R_{\rm p}^{\rm TH}$. As $R_{\rm p}^{\rm TH}$ increases, the WSSE of all SIB-RIS-assisted schemes decreases. This is because a more stringent PU QoS requirement forces the PT beamforming and SIB-RIS configuration to allocate more power and spatial DoFs to strengthening the primary link and suppressing the interference at the PU, leaving fewer resources for the secondary transmission. This behavior reflects the inherent tradeoff between PU's protection and SN throughput in underlay CR systems. Over the entire range of $R_{\rm p}^{\rm TH}$, the \textbf{Proposed} scheme maintains a clear WSSE advantage over the \textbf{SUD}, \textbf{TDMA}, and \textbf{AA-NOMA} benchmarks. In particular, the AA-NOMA curves exhibit a pronounced WSSE degradation as $R_{\rm p}^{\rm TH}$ grows, whereas the SIB-RIS-based schemes experience a much slower performance loss, highlighting the benefit of exploiting the SIB-RIS-based architecture.
	
	\section{Conclusion}
In this paper, we proposed an underlay CR framework integrating SIB-RISs with NOMA to enable spectrum-efficient secondary communications. By leveraging the PT's signal as both an RF carrier and energy source, the SIB-RIS-equipped STs achieve passive uplink NOMA transmission via power-splitting-based energy harvesting and information-bearing backscatter modulation, without requiring active RF chains or dedicated spectrum resources. To maximize the WSSE of the secondary network while ensuring PU's QoS, we developed an efficient BCD-based algorithm to jointly optimize the PT beamforming vector, the PS ratios, and the SIB-RIS reflection coefficients. Numerical results validate the efficacy of the proposed scheme, demonstrating substantial WSSE gains over benchmarks. Key insights reveal that increasing the PT's antenna count or SIB-RIS element scale enhances system performance. Future work may extend this framework to imperfect CSI scenarios, multi-PU environments, or hybrid active-passive RIS architectures to further advance green and massively connected 6G networks.
		
	\bibliographystyle{IEEEtran}
	\bibliography{Ref6}

@ARTICLE{R1,
  author={Al-Fuqaha, Ala and Guizani, Mohsen and Mohammadi, Mehdi and Aledhari, Mohammed and Ayyash, Moussa},
  journal={IEEE Commun. Surveys Tuts.}, 
  title={Internet of {T}hings: A Survey on Enabling Technologies, Protocols, and Applications}, 
  year={Fourthquarter 2015},
  volume={17},
  number={4},
  pages={2347-2376},
  doi={10.1109/COMST.2015.2444095}}

@ARTICLE{R2,
  author={Wang, Cheng-Xiang and others},
  journal={IEEE Commun. Surveys Tuts.}, 
  title={On the Road to 6{G}: Visions, Requirements, Key Technologies, and Testbeds}, 
  year={Secondquarter 2023},
  volume={25},
  number={2},
  pages={905-974},
  doi={10.1109/COMST.2023.3249835}}

@ARTICLE{R3,
  author={Mu, Xidong and Wang, Zhaolin and Liu, Yuanwei},
  journal={IEEE Wireless Commun.}, 
  title={{NOMA} for Integrating Sensing and Communications Toward 6{G}: A Multiple Access Perspective}, 
  year={2024},
  volume={31},
  number={3},
  pages={316-323},
  doi={10.1109/MWC.015.2200559}}

@ARTICLE{R4,
  author={Mahapatra, Rajarshi and Nijsure, Yogesh and Kaddoum, Georges and Ul Hassan, Naveed and Yuen, Chau},
  journal={IEEE Commun. Surveys Tuts.}, 
  title={Energy Efficiency Tradeoff Mechanism Towards Wireless Green Communication: A Survey}, 
  year={Firstquarter 2016},
  volume={18},
  number={1},
  pages={686-705},
  doi={10.1109/COMST.2015.2490540}}

@ARTICLE{R5,
  author={Wang, Huaxia and Yao, Yu-Dong},
  journal={IEEE Network}, 
  title={Secondary User Access Control in Cognitive Radio Networks: Concept, Design, and Analysis}, 
  year={2023},
  volume={37},
  number={6},
  pages={176-181},
  doi={10.1109/MNET.125.2200270}}

@article{R6,
	title = {Multiple-Antenna-Assisted Non-Orthogonal Multiple Access},
	author = {Liu, Yuanwei and Xing, Hong and Pan, Cunhua and Nallanathan, Arumugam and Elkashlan, Maged and Hanzo, Lajos},
	year = {2018},
month={Apr.},
	journal = {IEEE Wireless Commun.},
	volume = {25},
	number = {2},
	pages = {17--23},
	issn = {1536-1284},
	doi = {10.1109/MWC.2018.1700080},
	urldate = {2023-08-29},
	langid = {english}
}

@article{R7,
	title = {Exploiting Intelligent Reflecting Surfaces in {NOMA} Networks: Joint Beamforming Optimization},
	shorttitle = {Exploiting Intelligent Reflecting Surfaces in NOMA Networks},
	author = {Mu, Xidong and Liu, Yuanwei and Guo, Li and Lin, Jiaru and {Al-Dhahir}, Naofal},
	year = {2020},
month={Oct.},
	journal = {IEEE Trans. Wireless Commun.},
	volume = {19},
	number = {10},
	pages = {6884--6898},
	issn = {1536-1276, 1558-2248},
	doi = {10.1109/TWC.2020.3006915},
	langid = {english}
}

@ARTICLE{R8,
  author={Tu, Zhixing and Long, Ruizhe and Pei, Yiyang and Liang, Ying-Chang},
  journal={IEEE Trans. Wireless Commun.}, 
  title={{RIS}-Enabled Full-Duplex Backscatter Communication in Multi-User Symbiotic Radio}, 
  year={2024},
month={Nov.},
  volume={23},
  number={11},
  pages={16261-16274},
  doi={10.1109/TWC.2024.3439622}}

@ARTICLE{R80,
  author={Yuan, Jie and Liang, Ying-Chang and Joung, Jingon and Feng, Gang and Larsson, Erik G.},
  journal={IEEE Trans. Commun.}, 
  title={Intelligent Reflecting Surface-Assisted Cognitive Radio System}, 
  year={2021},
  month={Jan.},
  volume={69},
  number={1},
  pages={675-687},
  doi={10.1109/TCOMM.2020.3033006}}

@ARTICLE{R9,
	author={Gong, Shimin and others},
	journal={IEEE Commun. Surveys Tuts.}, 
	title={Toward Smart Wireless Communications via Intelligent Reflecting Surfaces: A Contemporary Survey}, 
	year={Fourthquarter 2020},
	volume={22},
	number={4},
	pages={2283-2314},
	doi={10.1109/COMST.2020.3004197}}

@article{R10,
	title = {Joint Beamforming Design and Power Splitting Optimization in {IRS}-Assisted {SWIPT} {NOMA} Networks},
	author = {Li, Zhendong and Chen, Wen and Wu, Qingqing and Wang, Kunlun and Li, Jun},
	year = {2022},
  month={Mar.},
	journal = {IEEE Trans. Wireless Commun.},
	volume = {21},
	number = {3},
	pages = {2019--2033},
	issn = {1536-1276, 1558-2248},
	doi = {10.1109/TWC.2021.3108901},
	langid = {english}
}

@ARTICLE{R11,
  author={Wan, Weidong and Liu, Yi and Zhang, Hailin},
  journal={IEEE Trans. Cogn. Commun. Netw.}, 
  title={Cooperative Beamforming and Flexible Index Modulation for Reconfigurable Intelligent Surface-Aided Symbiotic Radio Systems}, 
  year={2024},
month={Feb.},
  volume={10},
  number={1},
  pages={180-197},
  doi={10.1109/TCCN.2023.3312391}}

@article{R13,
	title = {Ambient Backscatter Communications: A Contemporary Survey},
	shorttitle = {Ambient Backscatter Communications},
	author = {Van Huynh, Nguyen and Hoang, Dinh Thai and Lu, Xiao and Niyato, Dusit and Wang, Ping and Kim, Dong In},
	year = {Fourthquarter 2018},
	journal = {IEEE Commun. Surveys Tuts.},
	volume = {20},
	number = {4},
	pages = {2889--2922},
	issn = {1553-877X},
	doi = {10.1109/COMST.2018.2841964},
	langid = {english}
}

@ARTICLE{R14,
  author={Bletsas, Aggelos and Alevizos, Panos N. and Vougioukas, Georgios},
  journal={IEEE Signal Process. Mag.}, 
  title={The Art of Signal Processing in Backscatter Radio for $\mu$W (or Less) {I}nternet of {T}hings: Intelligent Signal Processing and Backscatter Radio Enabling Batteryless Connectivity}, 
  year={2018},
month={Sep.},
  volume={35},
  number={5},
  pages={28-40},
  doi={10.1109/MSP.2018.2837678}}

@ARTICLE{R15,
  author={Hoang, Dinh Thai and Niyato, Dusit and Wang, Ping and Kim, Dong In and Han, Zhu},
  journal={IEEE Trans. Commun.}, 
  title={Ambient Backscatter: A New Approach to Improve Network Performance for {RF}-Powered Cognitive Radio Networks}, 
  year={2017},
month={Sep.},
  volume={65},
  number={9},
  pages={3659-3674},
  doi={10.1109/TCOMM.2017.2710338}}

@ARTICLE{R16,
  author={Liang, Ying-Chang and Zhang, Qianqian and Larsson, Erik G. and Li, Geoffrey Ye},
  journal={IEEE Trans. Cogn. Commun. Netw.}, 
  title={Symbiotic Radio: Cognitive Backscattering Communications for Future Wireless Networks}, 
  year={2020},
month={Dec.},
  volume={6},
  number={4},
  pages={1242-1255},
  doi={10.1109/TCCN.2020.3023139}}

@ARTICLE{R17,
  author={Xu, Yongjun and Jiang, Siqiao and Xue, Qing and Li, Xingwang and Yuen, Chau},
  journal={IEEE Internet Things J.}, 
  title={Throughput Maximization for {NOMA}-Based Cognitive Backscatter Communication Networks With Imperfect {CSI}}, 
  year={2023},
month={Nov.},
  volume={10},
  number={22},
  pages={19595-19606},
  doi={10.1109/JIOT.2023.3289181}}

@ARTICLE{R18,
  author={Guo, Huayan and Zhang, Qianqian and Xiao, Sa and Liang, Ying-Chang},
  journal={IEEE Internet Things J.}, 
  title={Exploiting Multiple Antennas for Cognitive Ambient Backscatter Communication}, 
  year={2019},
month={Feb.},
  volume={6},
  number={1},
  pages={765-775},
  doi={10.1109/JIOT.2018.2856633}}

@ARTICLE{R19,
  author={Khan, Wali Ullah and Li, Xingwang and Zeng, Ming and Dobre, Octavia A.},
  journal={IEEE Commun. Lett.}, 
  title={Backscatter-Enabled {NOMA} for Future 6{G} Systems: A New Optimization Framework Under Imperfect {SIC}}, 
  year={2021},
month={May},
  volume={25},
  number={5},
  pages={1669-1672},
  doi={10.1109/LCOMM.2021.3052936}}

@ARTICLE{R20,
  author={Valentini, Roberto and Di Marco, Piergiuseppe and Santucci, Fortunato},
  journal={IEEE Commun. Lett.}, 
  title={A {NOMA} Scheme for {IoT} Enabled by Selective Powering of Passive Backscattering Nodes}, 
  year={2022},
month={Sep.},
  volume={26},
  number={9},
  pages={2195-2199},
  doi={10.1109/LCOMM.2022.3187220}}

@article{R21,
	title = {Metasurface-Assisted Massive Backscatter Wireless Communication with Commodity {Wi-Fi} Signals},
	author = {Zhao, Hanting and Shuang, Ya and Wei, Menglin and Cui, Tie Jun and Hougne, Philipp Del and Li, Lianlin},
	year = {2020},
month={Aug.},
	journal = {Nat. Commun.},
	volume = {11},
	number = {1},
	pages = {3926},
	issn = {2041-1723},
	doi = {10.1038/s41467-020-17808-y},
	urldate = {2023-08-24},
	langid = {english}
}

@ARTICLE{R22,
  author={Zhou, Hu and Liang, Ying-Chang and Yuen, Chau and Niyato, Dusit},
  journal={IEEE Wireless Communications}, 
  title={What Can {RIS}s Do for Symbiotic Radios?}, 
  year={2025},
  volume={},
  number={},
  note    = {early access, DOI: 10.1109/MWC.2025.3601677},
  doi={10.1109/MWC.2025.3601677}}

@ARTICLE{R23,
  author={Han, Shuai and Wang, Jinming and Li, Cheng and Hossain, Ekram},
  journal={IEEE Commun. Surveys Tuts.}, 
  title={Principles, Applications, and Challenges of Reconfigurable Intelligent Surface-Enabled Backscatter Communication: A Comprehensive Survey and Outlook}, 
  year={2025},
 month={Oct.},
  volume={27},
  number={5},
  pages={2937-2972},
  doi={10.1109/COMST.2024.3519788}
}

@ARTICLE{R24,
  author={Liang, Ying-Chang and Zhang, Qianqian and Wang, Jun and Long, Ruizhe and Zhou, Hu and Yang, Gang},
  journal={Proc. IEEE}, 
  title={Backscatter Communication Assisted by Reconfigurable Intelligent Surfaces}, 
  year={2022},
month={Sep.},
  volume={110},
  number={9},
  pages={1339-1357},
  doi={10.1109/JPROC.2022.3169622}}

@article{R25,
	title = {Is Backscatter Link Stronger than Direct Link in Reconfigurable Intelligent Surface-Assisted System?},
	author = {Zhao, Wenjing and Wang, Gongpu and Atapattu, Saman and Tsiftsis, Theodoros A. and Tellambura, Chintha},
	year = {2020},
month={Jun.},
	journal = {IEEE Commun. Lett.},
	volume = {24},
	number = {6},
	pages = {1342--1346},
	issn = {1558-2558},
	doi = {10.1109/LCOMM.2020.2980510},
	langid = {english}
}

@article{R26,
	title = {{MIMO} Transmission Through Reconfigurable Intelligent Surface: System Design, Analysis, and Implementation},
	shorttitle = {MIMO Transmission Through Reconfigurable Intelligent Surface},
	author = {Tang, Wankai and Dai, Jun Yan and Chen, Ming Zheng and Wong, Kai-Kit and Li, Xiao and Zhao, Xinsheng and Jin, Shi and Cheng, Qiang and Cui, Tie Jun},
	year = {2020},
month={Nov.},
	journal = {IEEE J. Sel. Areas Commun.},
	volume = {38},
	number = {11},
	pages = {2683--2699},
	issn = {1558-0008},
	doi = {10.1109/JSAC.2020.3007055},
	langid = {english}
}

@ARTICLE{R27,
  author={Zhao, Yang and Clerckx, Bruno},
  journal={IEEE J. Sel. Areas Commun.}, 
  title={{RIS}catter: Unifying Backscatter Communication and Reconfigurable Intelligent Surface}, 
  year={2024},
month={Jun.},
  volume={42},
  number={6},
  pages={1642-1655},
  doi={10.1109/JSAC.2024.3389114}}

@article{R28,
	title = {Joint Spatial and Reflecting Modulation for Active {IRS} Assisted {MIMO} Communications},
	author = {Sanila, K. S. and Rajamohan, Neelakandan},
	year = {2023},
month={May},
	journal = {IEEE Trans. Commun.},
	volume = {71},
	number = {5},
	pages = {3132--3143},
	issn = {0090-6778, 1558-0857},
	doi = {10.1109/TCOMM.2023.3258486},
	urldate = {2023-08-31},
	langid = {english}
}

@article{R29,
	title = {Intelligent Reflecting Surface Backscatter Enabled Downlink Multi-Cell {MIMO} Networks},
	author = {Xu, Sai and Zhang, Jiliang and Liu, Jiajia and Du, Yanan and Zhang, Jie},
	year={2024},
month={Jan.},
        volume={23},
        number={1},
        pages={171-184},
	journal = {IEEE Trans. Wireless Commun.},
	issn = {1558-2248},
	doi = {10.1109/TWC.2023.3276484},
	langid = {english}
}

@article{R30,
	title = {Matrix-Calibration-Based Cascaded Channel Estimation for Reconfigurable Intelligent Surface Assisted Multiuser {MIMO}},
	author = {Liu, Hang and Yuan, Xiaojun and Zhang, Ying-Jun Angela},
	year = {2020},
month={Nov.},
	journal = {IEEE J. Sel. Areas Commun.},
	volume = {38},
	number = {11},
	pages = {2621--2636},
	issn = {1558-0008},
	doi = {10.1109/JSAC.2020.3007057},
	langid = {english}
}

@article{R31,
	title = {Channel Estimation for {RIS}-Empowered Multi-User {MISO} Wireless Communications},
	author = {Wei, Li and Huang, Chongwen and Alexandropoulos, George C. and Yuen, Chau and Zhang, Zhaoyang and Debbah, M{\'e}rouane},
	year = {2021},
month={Jun.},
	journal = {IEEE Trans. Commun.},
	volume = {69},
	number = {6},
	pages = {4144--4157},
	issn = {1558-0857},
	doi = {10.1109/TCOMM.2021.3063236},
	langid = {english}
}

@ARTICLE{R32,
  author={Lv, Lu and Luo, Hao and Li, Zan and Wu, Qingqing and Ding, Zhiguo and Al-Dhahir, Naofal and Chen, Jian},
  journal={IEEE Trans. Wireless Commun.}, 
  title={Self-Sustainable Intelligent Omni-Surface Aided Wireless Networks: Protocol Design and Resource Allocation}, 
  year={2024},
  month={Jul.},
  volume={23},
  number={7},
  pages={7503-7519},
  doi={10.1109/TWC.2023.3342037}}

@ARTICLE{R33,
  author={Zhu, Zhengyu and Li, Zheng and Chu, Zheng and Sun, Gangcan and Hao, Wanming and Liu, Peijia and Lee, Inkyu},
  journal={IEEE Trans. Wireless Commun.}, 
  title={Resource Allocation for Intelligent Reflecting Surface Assisted Wireless Powered {IoT} Systems With Power Splitting}, 
  year={2022},
  month={May},
  volume={21},
  number={5},
  pages={2987-2998},
  doi={10.1109/TWC.2021.3117346}}

@ARTICLE{R34,
  author={Zeng, Ming and Li, Xingwang and Li, Gen and Hao, Wanming and Dobre, Octavia A.},
  journal={IEEE Commun. Lett.}, 
  title={Sum Rate Maximization for {IRS}-Assisted Uplink {NOMA}}, 
  year={2021},
  month={Jan.},
  volume={25},
  number={1},
  pages={234-238},
  doi={10.1109/LCOMM.2020.3025978}}

@article{R35,
	title = {Fractional Programming for Communication Systems\textemdash Part {II}: Uplink Scheduling via Matching},
	shorttitle = {Fractional Programming for Communication Systems\textemdash Part II},
	author = {Shen, Kaiming and Yu, Wei},
	year = {2018},
  month={May},
	journal = {IEEE Trans. Signal Process.},
	volume = {66},
	number = {10},
	pages = {2631--2644},
	issn = {1053-587X, 1941-0476},
	doi = {10.1109/TSP.2018.2812748},
	urldate = {2023-02-20},
	langid = {english}
}

@article{Complexity,
	title = {Semidefinite Relaxation of Quadratic Optimization Problems},
	author = {Luo, Zhi-quan and Ma, Wing-kin and So, Anthony Man-cho and Ye, Yinyu and Zhang, Shuzhong},
	year = {2010},
month={May},
	journal = {{IEEE} Signal Processing Mag.},
	volume = {27},
	number = {3},
	pages = {20--34},
	issn = {1558-0792},
	doi = {10.1109/MSP.2010.936019},
	langid = {english}
}

@ARTICLE{Xu,
  author={Xu, Sai and Chen, Chen and Du, Yanan and Wang, Jiangzhou and Zhang, Jie},
  journal={IEEE Trans. Wireless Commun.}, 
  title={Intelligent Reflecting Surface Backscatter Enabled Uplink Coordinated Multi-Cell {MIMO} Network}, 
  year={2023},
  volume={22},
  number={8},
  pages={5685-5696},
  doi={10.1109/TWC.2023.3236405}}

@ARTICLE{Add1,
  author={Sun, Zeyang and Xu, Sai and Han, Shuai and Wang, Jinming and Wu, Chenyu and Li, Cheng},
  journal={IEEE Trans. Veh. Technol.}, 
  title={Information-Bearing {RIS} Assisted {NOMA} Communication for {MEC} Networks}, 
  year={2025},
  volume={},
  number={},
  note= {early access, DOI: 10.1109/TVT.2025.3621248},
  doi={10.1109/TVT.2025.3621248}}

@ARTICLE{Add2,
  author={Wu, Chenyu and You, Changsheng and Liu, Yuanwei and Chen, Li and Shi, Shuo},
  journal={IEEE Trans. Veh. Technol.}, 
  title={Two-Stage Hierarchical Beam Training for Near-Field Communications}, 
  year={2024},
  volume={73},
  number={2},
  pages={2032-2044},
  doi={10.1109/TVT.2023.3311868}}
			
	\appendices
	\section{Proof of Lemma 1}
	For fixed $\left(\boldsymbol{\alpha}, \boldsymbol{\beta}, \mathbf{w}, \mathcal{R}\right)$, 
	problem (P5) is convex with respect to the PS vector $\boldsymbol{\delta}$. 
	In particular, the objective in~\eqref{eq21a} can be rewritten as
	\begin{equation}
		f(\boldsymbol{\delta})
		= \sum_{j=1}^M a_j \delta_j
		- \sum_{i=1}^M c_i \delta_i^2,
	\end{equation}
	where
	\begin{equation}
		a_j =
		2\sqrt{\omega_j(1+\alpha_j)}\,
		{\Re}\!\left\{
		\beta_j^* \mathbf{g}_j^H 
		\mathrm{diag}(\boldsymbol{\phi}_j)\mathbf{F}_j\mathbf{w}
		\right\},
	\end{equation}
	\begin{equation}
		c_i =
		{\left| {{\bf{g}}_i^H{\rm{diag}}({{\bm \phi} _i}){{\bf{F}}_i}{\bf{w}}} \right|^2}
		\sum_{j=1}^{i} |\beta_j|^2 \;\ge 0,
	\end{equation}
	are constants independent of $\boldsymbol{\delta}$. 
	Hence, $f(\boldsymbol{\delta})$ is a separable quadratic function with diagonal Hessian 
	$\nabla^2 f(\boldsymbol{\delta}) = -2\,\mathrm{diag}(c_1,\ldots,c_M) \preceq \mathbf{0}$, 
	and is therefore concave in $\boldsymbol{\delta}$. Hence, $f(\boldsymbol{\delta})$ is concave in $\boldsymbol{\delta}$.

	The box constraints in~\eqref{eq7d} impose $0 < \delta_j < 1$, which define a convex set. 
	The EH constraints in~\eqref{eq7e} can be equivalently written as
	\begin{equation}
		\delta_j^2 
		\le 1 - \frac{\mu K}{\chi \left\|\mathbf{F}_j \mathbf{w}\right\|^2},
		\quad j \in \mathcal{M},
	\end{equation}
	which are convex quadratic inequalities in $\delta_j$. 
	Moreover, the PU's QoS constraint in~\eqref{eq7f} becomes
	\begin{equation}
		\sum_{j=1}^M k_j \delta_j^2 
		\le C,
	\end{equation}
	where $k_j = \big\|\mathbf{g}_{{\rm p}j}^H 
	\mathrm{diag}(\boldsymbol{\phi}_j)\mathbf{F}_j\mathbf{w}\big\|^2 \ge 0$ and 
	$C$ is a constant determined by $\mathbf{w}$ and system parameters. 
	This is again a convex quadratic constraint in $\boldsymbol{\delta}$. 
	
	Therefore, problem (P5) maximizes a concave objective over a convex feasible set, and thus the optimal PS vector $\boldsymbol{\delta}$ can be computed efficiently using convex solvers such as the CVX toolbox.

		\end{document}